\documentclass[a4paper,11pt]{article}
\usepackage[utf8]{inputenc}
\usepackage[english]{babel}
\usepackage{braket}
\usepackage{xspace}
\usepackage{bbm}
\usepackage{mathdots}
\usepackage{stackrel}
\usepackage[dvipsnames]{xcolor}
\usepackage{appendix}
\usepackage{hyperref}
\hypersetup{
 colorlinks=true,
 linkcolor=blue,
 anchorcolor = blue,
 citecolor = blue,
 filecolor = blue,
 urlcolor = blue
}
\usepackage{mathtools}
\usepackage{bbm}
\usepackage{comment}
\usepackage{subfigure}
\usepackage[T1]{fontenc}               
\usepackage[left=3cm,
            right=2.5cm,
            top=2.5cm,
            bottom=3cm
            ]{geometry}                
\usepackage{graphicx}
\usepackage{mathrsfs}
\usepackage{appendix}
\usepackage{fancyhdr}
\usepackage{amsmath,                   
            amssymb,                   
            amsthm}                    

\usepackage{setspace}
\usepackage{cite}
\usepackage{cancel}
\usepackage[margin=30pt, bf, font=small, center, justification=justified]{caption}[2004/07/16]

\usepackage{setspace}

\title{\bf Entanglement asymmetry and quantum Mpemba effect in the XY spin chain}

\author{Sara Murciano$^{1,2}$, Filiberto Ares$^3$, Israel Klich$^4$, and Pasquale Calabrese$^{3,5}$}

\date{}

\begin{document} 

\maketitle
{\small
\vspace{-5mm}  \ \\
$^{1}$	Walter Burke Institute for Theoretical Physics, Caltech, Pasadena, CA 91125, USA\\
\medskip
$^{2}$	Department of Physics and IQIM, Caltech, Pasadena, CA 91125, USA\\[-0.1cm]
\medskip
{$^{3}$}  SISSA and INFN Sezione di Trieste, via Bonomea 265, 34136 Trieste, Italy\\[-0.1cm]
\medskip
{$^{4}$}  Department of Physics, University of Virginia, Charlottesville, VA, USA\\[-0.1cm]
\medskip
{$^{5}$}  International Centre for Theoretical Physics (ICTP), Strada Costiera 11, 34151 Trieste, Italy\\[-0.1cm]
\medskip
}

\begin{abstract}
Entanglement asymmetry is a quantity recently introduced to measure how much a symmetry is broken in a part of an extended quantum system. It has been employed to analyze the non-equilibrium dynamics of a broken symmetry after a global quantum quench with a Hamiltonian that preserves it.   In this work, we carry out a comprehensive analysis of the entanglement asymmetry at equilibrium taking the ground state of the XY spin chain, which breaks the $U(1)$ particle number symmetry, and provide a physical interpretation of it in terms of superconducting Cooper pairs. We also consider quenches from this ground state to the XX spin chain, which preserves the $U(1)$ symmetry. In this case, the entanglement asymmetry reveals that the more the symmetry is initially broken, the faster it may be restored in a subsystem, a surprising and counter-intuitive phenomenon that is a type of a quantum Mpemba effect. We obtain a quasi-particle picture for the entanglement asymmetry in terms of Cooper pairs, from which we derive the microscopic conditions to observe the quantum Mpemba effect in this system, giving further support to the criteria recently proposed for arbitrary integrable quantum systems. In addition, we find that the power law governing symmetry restoration depends discontinuously on whether the initial state is critical or not, leading to new forms of strong and weak Mpemba effects.

\end{abstract}

 \tableofcontents

 \newpage
\section{Introduction}
Hot water may freeze faster than cold water: this counter-intuitive statement describes the Mpemba effect. Such phenomenon was already known to Aristotle and was neglected until 1963  when a student called E. B. Mpemba observed it preparing an ice-cream~\cite{cool}. This observation has opened a new research activity devoted to understanding the mechanism and conditions behind the Mpemba effect. Indeed, it has been observed not only in a solution of milk and sugar or in water but in a wide variety of systems, including clathrate hydrates~\cite{clathrate}, polymers~\cite{polymers}, magnetic alloys~\cite{magneto}, carbon nanotube resonators~\cite{carbon}, granular gases~\cite{granular}, or dilute atomic gases~\cite{cold} to cite some of them. Today, the Mpemba effect is more generally rephrased as an anomalous relaxation phenomenon where a system initially further out of equilibrium relaxes faster than a system initially closer to equilibrium. Recently, a theoretical framework for the Mpemba effect was developed in Ref.~\cite{raz,raz2}, followed by a demonstration of the effect in a controlled experimental setting consisting of a colloidal system that is suddenly quenched by placing it in a thermal bath at a lower temperature~\cite{kumar}. Further aspects of this framework have been studied, e.g., in \cite{wv-22,teza2023relaxation, wbv-23, bwv-23}. We emphasize that one important aspect of these works is the introduction of a distance between the state of the system and the final equilibrium state to characterize the Mpemba effect.

Despite considerable effort to understand this phenomenon at a classical level, there are only a few investigations in the quantum realm. Most of them study the relaxation of quantum systems after a quench of the temperature or are subject to non-unitary dynamics~\cite{quantum1,quantum2,quantum3,quantum4,quantum5,quantum6}. However, a version of the Mpemba effect in a closed many-body quantum system at zero temperature has been recently reported in Ref.~\cite{amc-22}. In particular, if we prepare a spin-$1/2$ chain in a state that breaks a $U(1)$ symmetry and we evolve the system unitarily with a Hamiltonian that preserves it, the symmetry may be dynamically restored in a subsystem of the chain and, furthermore, the more the symmetry is initially broken, the faster it may be restored. 
The situation here is slightly different from the standard classical Mpemba effect: since the system is isolated, the local equilibrium state does depend on the initial state. 
Then, what defines the quantum Mpemba effect in this context is not the distance from a common asymptotic state but rather the amount of symmetry breaking.

Thus, in order to study the quantum Mpemba effect, we have to use a quantity that does a similar job as the distance considered in Refs.~\cite{kumar, raz} to probe the classical counterpart, but at the level of symmetry breaking. To this aim 
in Ref.~\cite{amc-22}
the entanglement asymmetry was 
introduced to measure how much a symmetry is broken in a part of an extended quantum system. 
So far, the entanglement asymmetry has been studied for the $U(1)$ symmetry associated with transverse magnetization (particle number) in global quantum quenches to the XX spin chain from both the tilted ferromagnetic and Néel states, see Refs.~\cite{amc-22} and~\cite{amvc-23} respectively.
While in the first case, the quantum Mpemba effect can be observed, for the tilted Néel state the symmetry is not restored after the quench since the reduced density matrix relaxes to a non-Abelian Generalized Gibbs ensemble. In this case, the asymmetry tends at late times to different non-zero values depending on the initial state, and one cannot define a quantum Mpemba effect.
The entanglement asymmetry and the quantum Mpemba effect have also been analyzed in quenches from different initial states to interacting integrable Hamiltonians in the recent Ref.~\cite{abckmr-23}, in particular, the Lieb-Liniger model and the rule 54 quantum cellular 
automaton, using the space-time duality approach developed in Ref.~\cite{bkcccr-23}. 
In addition, a general explanation of the microscopic origin of the quantum Mpemba effect in free and interacting 
integrable systems has also been proposed in Ref.~\cite{abckmr-23}.  
Furthermore, experimental confirmations of this effect have been reported in a trapped-ion setup \cite{exp-ti}.
Entanglement asymmetry has also been employed to analyze the breaking of discrete symmetries in the XY spin-chain \cite{fac-23} and the massive Ising field theory~\cite{cm-23}, and of compact groups in matrix product states~\cite{cv-23}.

The goal of the present paper is twofold. On the one hand, we perform a comprehensive analysis
of the entanglement asymmetry in the ground state of the XY spin chain, which is the most paradigmatic
free integrable system that breaks a $U(1)$ symmetry. On the other hand, taking this ground
state, we investigate the time evolution of the entanglement asymmetry after a sudden global
quench to the XX spin chain Hamiltonian, which respects the $U(1)$ symmetry. This
framework provides the ideal setup to further study the quantum Mpemba effect discovered
in Ref.~\cite{amc-22} in free fermionic systems and give support to the general mechanism presented 
in Ref.~\cite{abckmr-23} for integrable models.

\textbf{Entanglement asymmetry}: Before summarizing our main results, let us first define the entanglement asymmetry. We consider an extended quantum system in a pure state $\ket{\Psi}$, 
which we divide into two spatial regions $A$ and $B$. The state of $A$ is given by the reduced density matrix $\rho_A$ 
obtained as $\rho_A=\mathrm{Tr}_B(\ket{\Psi}\bra{\Psi})$, where ${\rm Tr}_B$ denotes the partial trace in the subsystem $B$. Let us denote by $Q$ the charge operator with integer eigenvalues that generates 
a $U(1)$ symmetry group. We require that 
$Q$ is the sum of the charge in each region, $Q=Q_A+Q_B$. 
If $\ket{\Psi}$ has a defined charge, i.e. it is an eigenstate of $Q$, then it respects the corresponding symmetry and $[\rho_A, Q_A]=0$. The latter implies that $\rho_A$ is block-diagonal in the eigenbasis of $Q_A$ and each block corresponds to a particular charge sector. This situation has recently been intensively studied in the context of entanglement since entanglement entropy~\cite{klich08scaling,lr-14, goldstein, xavier} and other entanglement measures~\cite{cgs-18, mbc-21, czzc-21, dge-23} admit a decomposition in the charge sectors of the theory, which provides a much better understanding of numerous features of quantum many-body systems~\cite{brc-19, mgc-20, pbc-21-1, pbc-21-2, pvcc-22, bcckr-22, mac-23, lukin, azses, neven, vitale, rath}.


On the other hand, if $\ket{\Psi}$ is not eigenstate of $Q$, then it breaks the $U(1)$ symmetry generated by $Q$ and $[\rho_A, Q_A]\neq 0$. Therefore, $\rho_A$ is not block-diagonal. In this case, a proper measure of how much the symmetry is broken in the subsystem $A$ is the entanglement asymmetry, denoted by $\Delta S_{A}$, and defined as
\begin{equation}\label{eq:def}
 \Delta S_{A}=S(\rho_{A,Q})-S(\rho_A), \quad S(\rho)=-\mathrm{Tr}(\rho\log \rho).
\end{equation}
In this definition, the density matrix $\rho_{A, Q}$ is the result of projecting $\rho_A$ over all the charge sectors of $Q_A$; that is,  $\rho_{A,Q}=\sum_{q\in\mathbb{Z}}\Pi_q\rho_A\Pi_q$, where $\Pi_q$ denotes 
the projector onto the eigenspace of $Q_A$ with charge 
$q\in\mathbb{Z}$. The matrix $\rho_{A, Q}$ is therefore block-diagonal in the eigenbasis of $Q_A$. One can check that, due to the form of $\rho_{A, Q}$, the entanglement asymmetry $\Delta S_A$ is equal to the relative entropy 
between $\rho_A$ and $\rho_{A,Q}$, $\Delta 
S_{A}=\mathrm{Tr}[\rho_A(\log\rho_A-\log\rho_{A,Q})]$~\cite{mhms-22}. This identity implies that the entanglement asymmetry is non-negative, $\Delta S_A\geq 0$. The other important property as measure of symmetry breaking is that $\Delta S_A$ vanishes if and only if $[\rho_A, Q_A]=0$; that is, when the state of $A$ respects the symmetry associated to $Q_A$.

The entanglement asymmetry $\Delta S_{A}$ can be computed from the moments of the density matrices $\rho_A$ and $\rho_{A, Q}$ by applying the well-known replica trick for the entanglement entropy~\cite{hlw-94,cc-04}. If we define the Rényi entanglement asymmetry as
\begin{equation}\label{eq:replicatrick}
  \Delta S_{A}^{(n)}=S^{(n)}(\rho_{A, Q})-S^{(n)}(\rho_{A}),\quad  S^{(n)}(\rho)=\frac{1}{1-n}\log\mathrm{Tr}(\rho^n),
\end{equation}
one has that $\lim_{n\to 1}\Delta S_{A}^{(n)}=\Delta S_{A}$. 
As we will see, $\Delta S_A^{(n)}$ is easier to calculate for positive integer $n$ values, for which it can be measured in ion trap experiments using protocols based on randomized shadows~\cite{exp-ti,vermersch, brydges, huang, elben}. Moreover, $\Delta S_A^{(n)}$ satisfies the two crucial properties to be a measure of symmetry breaking: it is non-negative~\cite{hms-23} and is zero if and only if $[\rho_A, Q_A]=0$.

\textbf{Main results:} As we have already mentioned, the goal of this 
work is to expand the analysis done in~\cite{amc-22} for the tilted 
ferromagnetic state. We study the entanglement asymmetry in the ground 
state of the XY spin chain, described by the Hamiltonian 
\begin{equation}\label{eq:Ham_XY}
H=-\frac{1}{4}\sum_{j=-\infty}^{\infty}[(1+\gamma)\sigma^x_j \sigma^x_{j+1}+(1-\gamma)\sigma^y_j \sigma^y_{j+1} +2h \sigma_j^z],
\end{equation}
where the $\sigma_j^{\beta}$ are the  Pauli matrices at the site $j$, $\gamma$ is the anisotropy parameter between the couplings in the $x$ and $y$ directions of the spin and $h$ is the value of the external transverse magnetic field. When the anisotropy parameter $\gamma$ is not zero, $H$ breaks the $U(1)$ symmetry generated by the transverse 
magnetization
\begin{equation}\label{eq:transverese_mag}
Q=\frac{1}{2}\sum_j \sigma_j^z.
\end{equation}
Therefore, in this work, we are interested in the region $\gamma\neq 0$
for which the ground state entanglement asymmetry associated to $Q$
is non-zero while for $\gamma=0$ it vanishes. For $\gamma\neq 0$, the XY spin chain is critical along the lines $|h|=1$, which belong to the Ising universality class. The tilted ferromagnetic states considered in Ref.~\cite{amc-22} are only a subset of the ground states of the Hamiltonian~\eqref{eq:Ham_XY} along the curve $\gamma^2+h^2=1$~~\cite{ktm-82, ms-85}.  In this more general setup, we can compute the entanglement asymmetry for any ground state of~\eqref{eq:Ham_XY} and we find that, for a subsystem $A$ of contiguous spins of length $\ell$, it reads 
\begin{equation}\label{eq:main_Result_gs}
\Delta S_A^{(n)}=\frac{1}{2}\log \ell+\frac{1}{2} \log \frac{\pi g(\gamma,h)n^{1/(n-1)}}{4}+O(\ell^{-1}),
\end{equation}
where $g(\gamma,h)$ is a function depending on the two parameters $\gamma$ and $h$ of the Hamiltonian~\eqref{eq:Ham_XY}. We find that this term is related to the density of Cooper pairs, which are responsible for the breaking of the conservation of the number of particles. 
We remark that $\Delta S_A^{(n)}$ increases logarithmically with the subsystem size, both if the system is critical ($|h|=1$) or not. 

If we choose the ground state of the Hamiltonian in Eq.~\eqref{eq:Ham_XY} for arbitrary $\gamma$ and $h$ and we let it evolve with the XX spin chain, that is taking $\gamma=0$ and $h=0$ in Eq.~\eqref{eq:Ham_XY}, which commutes with the charge~\eqref{eq:transverese_mag}, the symmetry is dynamically restored. We derive a quasi-particle picture for the entanglement asymmetry at large times after the quench based on the initial density of Cooper pairs. From it, we find that the R\'enyi entanglement asymmetry vanishes for large times as $t^{-3}$ for any initial value of $\gamma$ when $|h|\neq 1$ and we predict under which conditions for the parameters $(\gamma,h)$ we observe the Mpemba effect. It turns out that, if the density of Cooper pairs around the slowest modes of the post-quench Hamiltonian is larger for the state that initially breaks less the symmetry, the quantum Mpemba effect occurs, in agreement with the general findings of~\cite{abckmr-23} for integrable systems. 
On the other hand, when the system is prepared initially in the critical ground state, i.e. $|h|=1$, the R\'enyi asymmetry vanishes as $t^{-1}$ for any value of $\gamma$. Therefore, for critical systems, we can define a \textit{strong} version of the Mpemba effect for which the relaxation happens algebraically slower regardless of the initial condition for the non-critical state. 

\textbf{Outline:} In Section~\ref{sec:charged_xy}, we provide a recipe to evaluate the entanglement asymmetry for Gaussian fermionic operators such as the reduced density matrix of the ground state of the XY spin chain. Section~\ref{sec:xy_gs} is devoted to the analysis of the entanglement asymmetry in the ground state of the XY model, while Section~\ref{sec:quench} studies the time-evolution of the asymmetry after a quench to the XX spin chain and the origin of the Mpemba effect. Finally, we draw our conclusions in Section~\ref{sec:conclusions} and we include three appendices with additional results and technical details.

\section{Charged moments and XY spin chain}\label{sec:charged_xy}

As we have seen in the previous section, by applying the replica trick, the entanglement asymmetry $\Delta S_{A}$ can be computed from the R\'enyi version $\Delta S^{(n)}_{A}$ defined in Eq.~\eqref{eq:replicatrick}. The advantage of doing this is that, using the Fourier
representation of the projector $\Pi_q$, the projected density matrix $\rho_{A, Q}$ can be rewritten as 
\begin{equation}
 \rho_{A, Q}=\int_{-\pi}^\pi \frac{{\rm d}\alpha}{2\pi}e^{-i\alpha Q_A}\rho_A e^{i\alpha Q_A}.
\end{equation}
Therefore, its moments are given by
\begin{equation}\label{eq:FT}
 \mathrm{Tr}(\rho_{A, Q}^n)=\int_{-\pi}^\pi \frac{{\rm d}\alpha_1\dots{\rm d}\alpha_n}{(2\pi)^n} Z_n(\boldsymbol{\alpha}),
 \end{equation}
where $\boldsymbol{\alpha}=\{\alpha_1,\dots,\alpha_n\}$ and $Z_n(\boldsymbol{\alpha})$ are the (generalized) charged moments
\begin{equation}\label{eq:Znalpha}
 Z_n(\boldsymbol{\alpha})=
 \mathrm{Tr}\left[\prod_{j=1}^n\rho_A e^{i\alpha_{j,j+1}Q_A}\right],
\end{equation}
with $\alpha_{ij}\equiv\alpha_i-\alpha_j$ and $\alpha_{n+1}=\alpha_1$. Since in general $[\rho_A, Q_A]\neq 0$, the order in which these operators enter in the expression of $Z_n(\boldsymbol{\alpha})$ is crucial. In fact, if $[\rho_A, Q_A]=0$, then $Z_n(\boldsymbol{\alpha})=Z_n(\boldsymbol{0})$, which implies $\mathrm{Tr}(\rho_{A, Q}^n)=\mathrm{Tr}(\rho_A^n)$
and $\Delta S_{A}^{(n)}=0$.


In this manuscript, we are particularly interested in calculating the 
charged moments $Z_n(\boldsymbol{\alpha})$ and, from them using Eq.~\eqref{eq:FT}, 
the Rényi entanglement asymmetry $\Delta S_A^{(n)}$ in the ground state
of the XY spin chain~\eqref{eq:Ham_XY}. As well-known, this Hamiltonian is easily diagonalizable as follows~\cite{lieb}. We can first map it to the fermionic operators $\boldsymbol{c}_j=(c_j^\dagger, c_j)$ via a Jordan-Wigner transformation, namely
\begin{equation}
H=-\frac{1}{2}\sum_{j=-\infty}^{\infty}\left(c^{\dagger}_j c_{j+1}+\gamma c^{\dagger}_jc^{\dagger}_{j+1} +\mathrm{h.c.}+2h c^{\dagger}_jc_j\right).
\end{equation}
By performing now a Fourier transformation to momentum space 
$d_k=\sum_{j\in\mathbb{Z}} e^{-ikj} c_j$
and then the Bogoliubov transformation
\begin{equation}
\left(\begin{array}{c}
 \eta_k \\
 \eta_{2\pi-k}^\dagger
 \end{array}\right)=
 \left(\begin{array}{cc}
  \cos(\Delta_k/2) & i \sin(\Delta_k/2) \\
  i\sin(\Delta_k/2) & \cos(\Delta_k/2)
  \end{array}
  \right)\left(\begin{array}{c} d_k \\ d_{2\pi-k}^\dagger\end{array}\right),
\end{equation}
with 
\begin{equation}
\begin{split}\label{eq:cs}
 \cos\Delta_k&=\frac{h-\cos(k)}{\sqrt{(h-\cos(k))^2+\gamma^2\sin^2 k}}, \\
  \sin\Delta_k&=\frac{\gamma \sin(k)}{\sqrt{(h-\cos(k))^2+\gamma^2\sin^2 k}},
 \end{split}
\end{equation}
the XY spin chain is diagonal in terms of the Bogoliubov modes $\eta_k$,
\begin{equation}
H=\sum_k \epsilon_k \left(\eta_k^\dagger \eta_k -\frac{1}{2}\right),
\end{equation}
where $\epsilon_k$ is the single-particle dispersion relation
\begin{equation}
\epsilon_k= \sqrt{(h-\cos k)^2+\gamma^2\sin^2 k}.
\end{equation}

Thus the ground state is the Bogoliubov vacuum   $\ket{0}$ that is  
annihilated by all the operators $\eta_k$, i.e. $\eta_k\ket{0}=0$ for all $k$. For $\gamma\neq 0$, this state breaks the $U(1)$ symmetry associated to the conservation of the total transverse magnetization~\eqref{eq:transverese_mag}, i.e. $[\rho, Q]\neq 0$ with $\rho=\ket{0}\bra{0}$, and the asymmetry $\Delta S_A^{(n)}$ is non-zero. On the other hand, for $\gamma=0$, $\ket{0}$ is an eigenstate of $Q$ and $\Delta S_A^{(n)}$ vanishes. Therefore, $\ket{0}$ is an ideal state to explore $\Delta S_{A}^{(n)}$.

The ground state of the XY spin chain is a Slater determinant and, consequently, the reduced density matrix $\rho_A$  is
a Gaussian operator in terms of $\boldsymbol{c}_j$~\cite{p-03}. 
This simplifies the calculation
of $\Delta S_{A}^{(n)}$ since, due to the Wick's theorem, $\rho_A$ is univocally determined by the 
two-point correlation matrix 
\begin{equation} \Gamma_{jj'}=2\mathrm{Tr}\left[\rho_A\boldsymbol{c}_j^\dagger
 \boldsymbol{c}_{j'}\right]-\delta_{jj'},
\end{equation}
with $j,j'\in A$. If $A$ is an interval of contiguous sites of length $\ell$, then $\Gamma$
is a $2\ell\times 2\ell$ block Toeplitz matrix; that is, their entries are the Fourier coefficients~\cite{FC-08}
\begin{equation}\label{eq:Gammat0}
\Gamma_{jj'}=\int_{0}^{2\pi}\frac{{\rm d}k}{2\pi}\mathcal{G}(k)e^{-ik(j-j')}, \quad j,j'=1,\dots,\ell,
\end{equation}
of the $2\times 2$ symbol
\begin{equation}\label{eq:gs_symbol}
    \mathcal{G}(k)=\left(\begin{array}{cc}
    \cos\Delta_k & -i \sin\Delta_k\\
     i\sin\Delta_k& -\cos\Delta_k
    \end{array}\right).
\end{equation}

Under the Jordan-Wigner transformation, the transverse magnetization $Q$ in Eq.~\eqref{eq:transverese_mag}
is mapped to the fermion number operator $Q=\sum_j (c_j^\dagger c_j -1/2)$ and  $e^{i\alpha Q_A}$ turns out to be Gaussian, too.
Therefore, Eq.~\eqref{eq:Znalpha} is the trace of the product of  Gaussian fermionic operators, $\rho_A$ and $e^{i\alpha_{j,j+1} Q_A}$. 
As explicitly shown in Appendix B of Ref.~\cite{amvc-23}, using the special properties of Gaussian operators~\cite{balian, FC10},
the trace of Eq.~\eqref{eq:Znalpha} can be re-expressed as a determinant involving the two-point correlation matrix $\Gamma$,
\begin{equation}\label{eq:numerics}
  Z_n(\boldsymbol{\alpha})=\sqrt{\det\left[\left(\frac{I-\Gamma}{2}\right)^n
  \left(I+\prod_{j=1}^n W_j\right)\right]},
\end{equation}
with $W_j=(I+\Gamma)(I-\Gamma)^{-1}e^{i\alpha_{j,j+1} n_A}$ and $n_A$ is a diagonal matrix with $(n_A)_{2j,2j}=1$, $(n_A)_{2j-1,2j-1}=-1$, $j=1, \cdots, \ell$. 
Eq.~\eqref{eq:numerics} allows to exactly compute numerically $\Delta S_A^{(n)}$ and is the starting point to derive analytic expressions for $Z_n(\boldsymbol{\alpha})$ and $\Delta S_A^{(n)}$ for large subsystem sizes.


\section{Entanglement asymmetry in the ground state of the XY spin chain}\label{sec:xy_gs}

In this section, we study the entanglement asymmetry in
the ground state of the XY spin chain. As we have previously shown, this state is the 
vacuum $\ket{0}$ of the Bogoliubov modes that diagonalize the 
Hamiltonian~\eqref{eq:Ham_XY} of the chain. Since the reduced density 
matrix $\rho_A={\rm Tr}_B(\ket{0}\bra{0})$ is Gaussian, we 
can apply Eq.~\eqref{eq:numerics} to study both numerically and 
analytically the charged moments 
$Z_n(\boldsymbol{\alpha})$, from which the R\'enyi 
entanglement asymmetry can be derived using Eqs.~\eqref{eq:FT} and~\eqref{eq:replicatrick}. 

\subsection{Charged moments}

\begin{figure}[t]
\centering
\subfigure
{\includegraphics[width=0.48\textwidth]{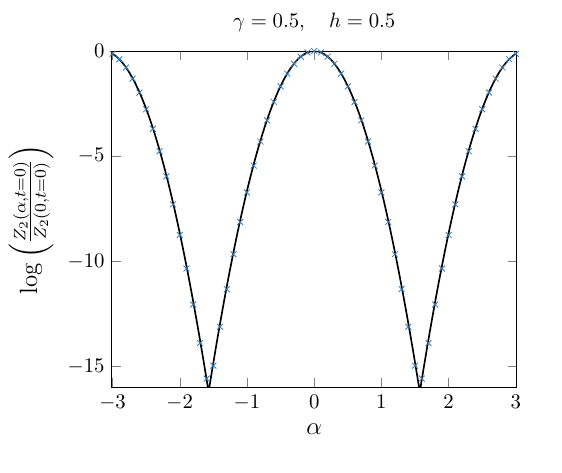}}
\subfigure
{\includegraphics[width=0.48\textwidth]{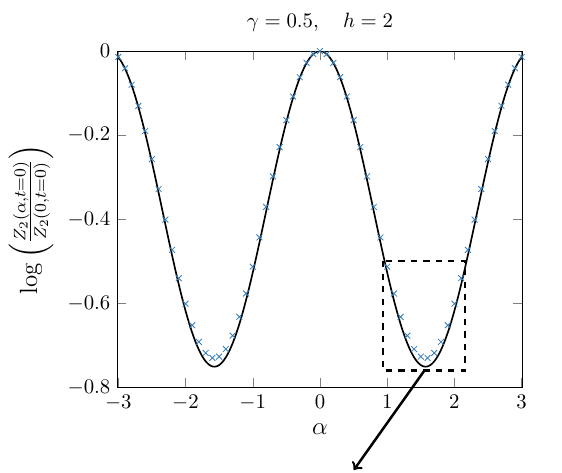}}
\subfigure
{\vspace{-2cm}\hspace{1.5cm}\includegraphics[width=0.48\textwidth]{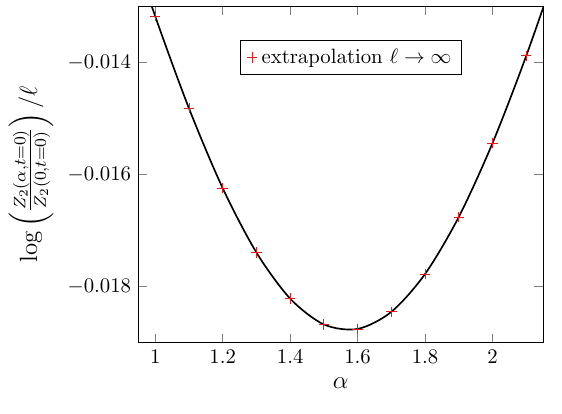}}
\caption{Logarithm of the $n=2$ charged moment $Z_2(\alpha)/Z_2(0)$ in the ground state of the XY spin chain as a function of $\alpha$ for different values of $h$ and $\gamma$ and subsystem size $\ell=40$.
The solid lines correspond to the analytic prediction of Eq.~\eqref{eq:Balpha_inftyt0} while the points are the exact values obtained using directly Eq.~\eqref{eq:charged_mom_n_2}.  The bottom panel represents the extrapolated data with extrapolation form $a+b/\ell$ to check that the discrepancy observed in the right panel is only a finite-size effect. To extrapolate the data, we have taken into account the numerical data for $\ell=40,50,60$. Interestingly, the extrapolated points are exactly equal to our analytical prediction of the linear growth of $\log Z_2(\alpha)$ in Eq.~\eqref{eq:Balpha_inftyt0} (solid curve). }\label{fig:eqfact}
\end{figure}

For simplicity, let us first consider the
case $n=2$ and afterwards we will generalize the results to any $n$.
Observe that, for $n=2$, the expression~\eqref{eq:numerics} of the charged moments 
$Z_n(\boldsymbol{\alpha})$ in terms of the two-point correlation function $\Gamma$
simplifies, after a change of variable $\alpha_{12}=\alpha$, as
\begin{equation}\label{eq:charged_mom_n_2}
Z_2(\alpha)=\sqrt{\det\left(\frac{I+\Gamma_\alpha\Gamma_{-\alpha}}{2}\right)}.
\end{equation}
The matrix $\Gamma_\alpha\equiv \Gamma e^{i\alpha n_A}$ is block Toeplitz 
\begin{equation}
(\Gamma_\alpha)_{jj'}=\int_{-\pi}^\pi \frac{{\rm d}k}{2\pi} \mathcal{G}_\alpha(k) e^{-ik(j-j')},\quad j, j'=1,\dots, \ell,
\end{equation}
with symbol
\begin{equation}
\mathcal{G}_{\alpha}(k)= 
\left( \begin{array}{cc}
e^{i\alpha} \cos\Delta_k & -i e^{-i\alpha} \sin\Delta_k\\
ie^{i\alpha}\sin\Delta_k & -e^{-i\alpha}\cos\Delta_k
\end{array}\right).
\end{equation}
Therefore, in Eq.~\eqref{eq:charged_mom_n_2}, we have the product $\Gamma_\alpha \Gamma_{-\alpha}$ of two block Toeplitz matrices, which in general is not block Toeplitz, and the well-known results on the  determinant of this kind of matrices cannot in principle be applied. However, in Ref.~\cite{amvc-23}, we found the following result for the asymptotic behavior of determinants that contain a product of block Toeplitz matrices like the one in Eq.~\eqref{eq:charged_mom_n_2}. If we denote as $T_\ell[g]$ the $(\ell d)\times (\ell d)$ dimensional block Toeplitz matrix with symbol the $d\times d$ matrix $g$, then for large $\ell$
\begin{equation}\label{eq:conj_1}
 \det\left(I+\prod_{j=1}^nT_\ell[g_j]\right)\sim e^{\ell A},
\end{equation}
where the coefficient $A$ is given by
\begin{equation}
A=\int_{0}^{2\pi}\frac{{\rm d}k}{2\pi}\log\det\left[I+\prod_{j=1}^n g_j(k)\right].
\end{equation}
If we apply Eq.~\eqref{eq:conj_1} in Eq.~\eqref{eq:charged_mom_n_2}, then we obtain that the $n=2$
charged moments behave for large subsystem size $\ell$ as
\begin{equation}\label{eq:asymp_charged_mom_n_2}
Z_2(\alpha) \sim Z_2(0) e^{A_2(\alpha)\ell},
\end{equation}
and
\begin{equation}\label{eq:A_2}
A_2(\alpha)=\int_{-\pi}^{\pi} \frac{{\rm d}k}{4\pi} \log(1-\sin^2\alpha\sin^2\Delta_k).
\end{equation}
In Fig.~\ref{fig:eqfact}, we numerically test this result. We plot the logarithm of the ground state charged moment $Z_2(\alpha)/Z_2(0)$ as a function of the angle $\alpha$ for a fixed subsystem of length $\ell=40$ and two different sets of values for $h$ and $\gamma$; in the left panel, we consider $h=\gamma=0.5$ while in the right one we take $h=2$ and $\gamma=0.5$. The dots are the exact value of $Z_2(\alpha)$ calculated using Eq.~\eqref{eq:charged_mom_n_2} and the solid lines correspond to the asymptotic analytic prediction of Eq.~\eqref{eq:asymp_charged_mom_n_2}. As evident in the plot, for $|h|\leq 1$, $\log (Z_2(\alpha)/Z_2(0))$
presents a cusp at $\alpha=\pm \pi/2$ while, for $|h|>1$, this non-analiticity disappears. In the inset of the right panel, we check that the discrepancy between the analytic prediction and the exact points around $\alpha=\pi/2$ is due to subleading corrections in $\ell$, see the caption for details.

\begin{figure}[t]
\centering
\subfigure
{\includegraphics[width=0.48\textwidth]{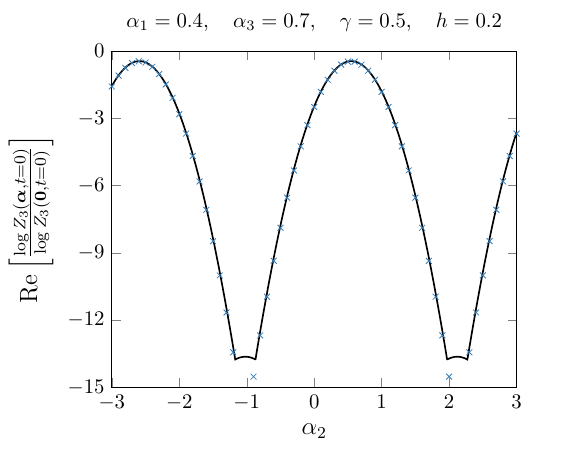}}
\subfigure
{\includegraphics[width=0.48\textwidth]{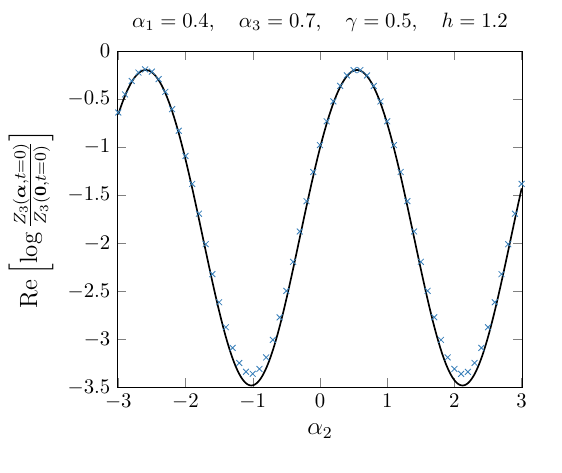}}
\subfigure
{\includegraphics[width=0.48\textwidth]{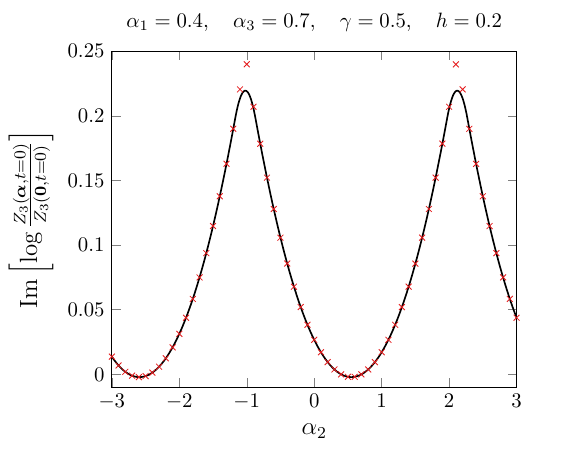}}
\subfigure
{\includegraphics[width=0.48\textwidth]{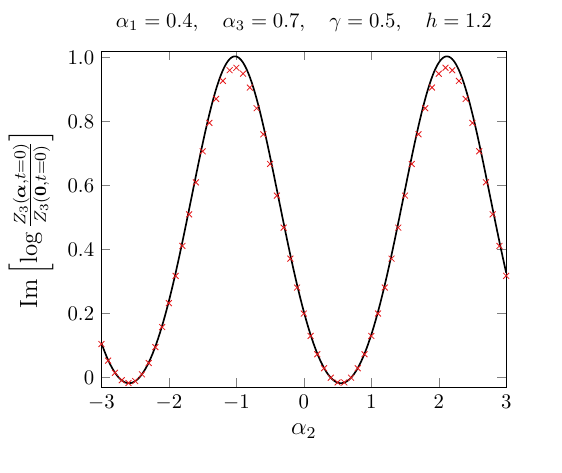}}
\caption{Logarithm of the $n=3$ charged moment $Z_3(\alpha_1,\alpha_2,\alpha_3)/Z_3(0,0,0)$ in the ground state of the XY spin chain  as a function of $\alpha_2$ for $\alpha_1$ and $\alpha_3$ constant, 
different values of the couplings $h$ and $\gamma$ and subsystem size $\ell=40$. In the upper
panels, we take its real part while the corresponding imaginary part is represented in the plots of the lower row. The points correspond to the exact numerical values calculated using Eq.~\eqref{eq:numerics}. 
The solid lines correspond to the asymptotic expression of Eq.~\eqref{eq:Balpha_inftyt0} employing as coefficient $A_3(\boldsymbol{\alpha})$ the prediction of Eq.~\eqref{eq:A_n}. }\label{fig:eqfactn3}
\end{figure}

The result of Eq.~\eqref{eq:asymp_charged_mom_n_2} for $n=2$ can be rewritten in a more appealing form that straightforwardly suggests its generalization to any integer $n\geq 2$. In fact, observe that the coefficient $A_2(\alpha)$ of Eq.~\eqref{eq:A_2} can be recast in the following factorized expression
\begin{equation}
    A_2(\alpha)=\int_{-\pi}^{\pi}\frac{{\rm d}k}{4\pi}\log(f(\cos\Delta_k, \alpha)f(\cos\Delta_k, -\alpha))
\end{equation}
where 
\begin{equation}\label{eq:f}
    f(\lambda,\alpha)=i\lambda \sin\left(\alpha\right)+\cos\left( \alpha\right).
\end{equation}
As we show in Appendix~\ref{app:finite_temperature}, this result can be extended to any integer $n\geq 2$. The charged moments behave for large $\ell$ similarly to the case $n=2$,
cf. Eq.~\eqref{eq:asymp_charged_mom_n_2},
\begin{equation}\label{eq:Balpha_inftyt0}
Z_n(\boldsymbol{\alpha})\sim Z_n(\boldsymbol{0})e^{\ell A_n(\boldsymbol{\alpha})},
\end{equation}
where the coefficient $A_n(\boldsymbol{\alpha})$ admits the following factorization in the replica space, 
\begin{equation}\label{eq:A_n}
A_n(\boldsymbol{\alpha})=\int_0^{2\pi}\frac{\mathrm{d}k}{4\pi}\log \prod_{j=1}^n f(\cos\Delta_k,\alpha_{j,j+1}).
\end{equation}
In Fig.~\ref{fig:eqfactn3}, we check numerically Eq.~\eqref{eq:Balpha_inftyt0} for the case $n=3$. \\
We consider the ratio
$Z_3(\alpha_1, \alpha_2, \alpha_3)/Z_3(0, 0, 0)$ as a function of $\alpha_2$ for given values of $\alpha_1$ and $\alpha_3$. Its real and imaginary parts are plotted respectively in the upper and lower panels for two different sets of couplings $h$ and $\gamma$: $h=0.2$, $\gamma=0.5$ on the left and $h=1.2$ and $\gamma=0.5$ on the right. We obtain an excellent agreement. As in the case $n=2$, the logarithm of $Z_3(\alpha_1, \alpha_2, \alpha_3)/Z_3(0, 0, 0)$ presents cusps when $|h|\leq 1$ that disappear in the phase $|h|>1$.

It is important to remark that, along the critical lines $|h|=1$, we have numerically observed that the expression~\eqref{eq:Znalpha} for the charged moments $Z_n(\boldsymbol{\alpha})$ includes an additional subleading term $Z_n(\boldsymbol{\alpha})\sim Z_n(\boldsymbol{0})e^{A_n(\boldsymbol{\alpha})\ell}\ell^{m_n(\boldsymbol{\alpha})}$. Unfortunately, the explicit form of $m_n(\boldsymbol{\alpha})$ cannot be obtained with the methods employed in this manuscript. However, since the factor $\ell^{m_n(\boldsymbol{\alpha})}$ produces a subleading term in the entanglement asymmetry, we can safely neglect it in the rest of the paper. 

\subsection{Asymptotic behavior of the entanglement asymmetry}

As we explain in Sec.~\ref{sec:charged_xy}, once we have the charged moments~\eqref{eq:Znalpha}, the Rényi entanglement asymmetry $\Delta S_A^{(n)}$ can be determined by plugging them into the the $n$-dimensional integral of Eq.~\eqref{eq:FT} and then using Eq.~\eqref{eq:replicatrick}. In general, this integral can only be calculated by numerical means but, employing a saddle point approximation, we can derive analytically the asymptotic behavior of $\Delta S_A^{(n)}$ for large subsystems.

To do so, we can follow the same strategy applied in Ref.~\cite{amvc-23}. By taking into account that the phases $\alpha_{jj+1}$ satisfy $\sum_{j=1}^n\alpha_{jj+1}=0$, we can reduce the $n$-fold integral~\eqref{eq:FT} to an $(n-1)$-fold one after the change of variables
$\tilde{\alpha}_j=\alpha_{jj+1}$,
\begin{equation}
{\rm Tr}(\rho_{A,Q}^n)=\int_{-\pi}^\pi\frac{{\rm d}\tilde{\alpha}_1\dots {\rm d}\tilde{\alpha}_{n-1}}{(2\pi)^{n-1}}{\rm Tr}\left(\rho_A e^{i\tilde{\alpha}_1 Q_A}\rho_A e^{i\tilde{\alpha}_2 Q_A}\cdots\ \rho_A e^{-i\sum_{j=1}^{n-1}\tilde{\alpha}_j Q_A}\right).
\end{equation}
If we insert in this expression the prediction of Eq.~\eqref{eq:Balpha_inftyt0} for the charged moments at large $\ell$, the integral takes the form
\begin{equation}\label{eq:int_1}
 \frac{{\rm Tr}(\rho_{A, Q}^n)}{{\rm Tr}(\rho_A^n)}
 \sim \int_{-\pi}^\pi \frac{{\rm d}\tilde{\alpha}_1\cdots {\rm d}\tilde{\alpha}_{n-1}}{(2\pi)^{n-1}} e^{\ell\left[\sum_{j=1}^{n-1} A_1(\tilde{\alpha}_j)+A_1(-\sum_{j=1}^{n-1}\tilde{\alpha}_j)\right]},
\end{equation}
where we have explicitly used the factorization in the replica space found in Eq.~\eqref{eq:A_n} for the coefficient $A_n(\boldsymbol{\alpha})$. One can check that there are $2^{n-1}$ points in the region of integration $[-\pi,\pi]^{\times(n-1)}$ that satisfy the saddle point condition
\begin{equation}
\partial_{\tilde{\alpha_i}}\left[\sum_{j=1}^{n-1} A_1(\tilde{\alpha}_j)+A_1\left(-\sum_{j=1}^{n-1}\tilde{\alpha}_j\right)\right]=0. 
\end{equation}
Around all the saddle points, the integrand of Eq.~\eqref{eq:int_1} has the same behavior at quadratic order in $\tilde{\alpha}_j$ and, therefore, their leading contribution to the integral is the same. Hence, if we expand the exponent in Eq.~\eqref{eq:int_1} around $\tilde{\alpha}_j=0$ and we properly count the number of saddle points, then Eq.~\eqref{eq:int_1} can be approximated by the Gaussian integral
\begin{equation}\label{eq:saddle_point_int}
\frac{{\rm Tr}(\rho_{A, Q}^n)}{{\rm Tr}(\rho_A^n)}\sim 2^{n-1}
\int_{-\infty}^\infty \frac{{\rm d}\tilde{\alpha}_1\cdots {\rm d}\tilde{\alpha}_{n-1}}{(2\pi)^{n-1}}
e^{-\frac{\ell g(\gamma, h)}{2}\left(\sum_{j=1}^{n-1}\tilde{\alpha}_j^2+\sum_{j<j'}\tilde{\alpha}_j\tilde{\alpha}_{j'}\right)},
\end{equation}
with
\begin{equation}\label{eq:g_gs}
g(\gamma,h)=\int_0^{2\pi}\frac{{\rm d}k}{2\pi} \sin^2\Delta_k.
\end{equation}
The integral of Eq.~\eqref{eq:saddle_point_int} is solvable using the standard formulae,
\begin{equation}\label{eq:spfactn}
\frac{\mathrm{Tr}(\rho_{A,Q}^n)}{\mathrm{Tr}(\rho_A^n)}=\frac{2^{n-1} }{(\pi \ell g(\gamma,h))^{(n-1)/2}n^{1/2}}+O(\ell^{-{(n+1)}/2}).
\end{equation}
Finally, plugging this result in Eq.~\eqref{eq:replicatrick}, we obtain that for the ground state of the XY spin chain, the Rényi entanglement asymmetry behaves as
\begin{equation}\label{eq:Deltaspn}
\Delta S_A^{(n)}=\frac{1}{2}\log \ell+\frac{1}{2} \log \frac{\pi g(\gamma,h)n^{1/(n-1)}}{4}+O(\ell^{-1}).
\end{equation}
The integral of Eq.~\eqref{eq:g_gs} that gives the term $g(\gamma, h)$ can be computed explicitly. In fact, if we perform the change of variables $z=e^{ik}$, it can be rewritten as a contour integral in the complex $z$-plane. Using then the residue theorem, we find
\begin{equation}\label{eq:g_gs_explicity}
 g(\gamma, h)=\begin{cases}
               \frac{\gamma}{\gamma+1},\quad &|h|\leq1, \\
               \frac{\gamma^2}{1-\gamma^2}\left(\frac{|h|}{\sqrt{h^2+\gamma^2-1}}-1\right), \quad& |h|>1.
              \end{cases}
\end{equation}

\begin{figure}[t]
\centering
\subfigure
{\includegraphics[width=0.49\textwidth]{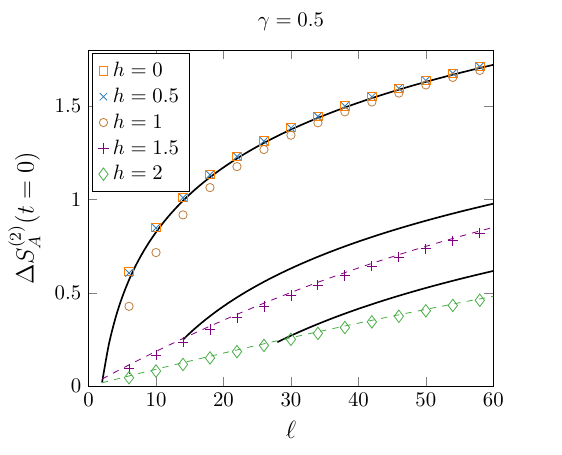}}
\subfigure
{\includegraphics[width=0.49\textwidth]{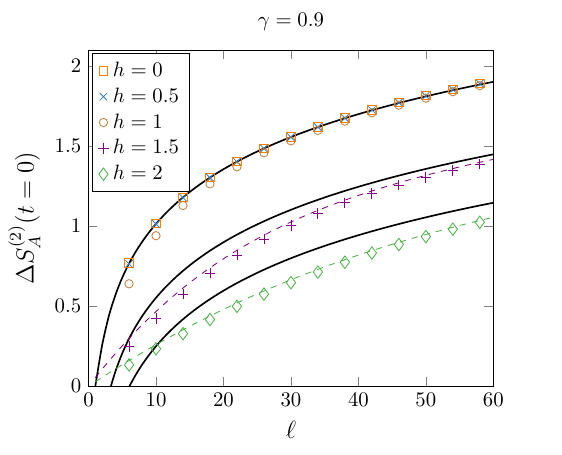}}
\subfigure
{\includegraphics[width=0.49\textwidth]{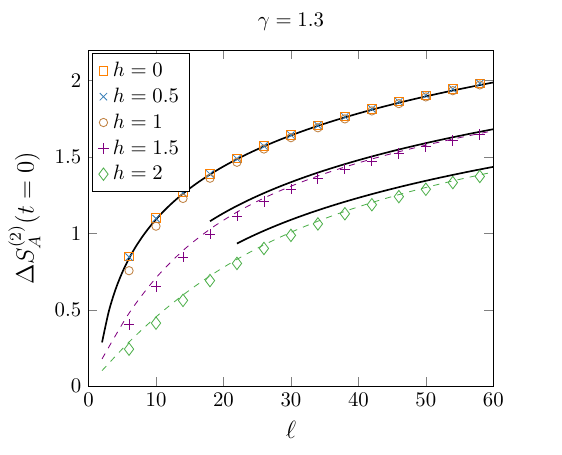}}
\subfigure
{\includegraphics[width=0.49\textwidth]{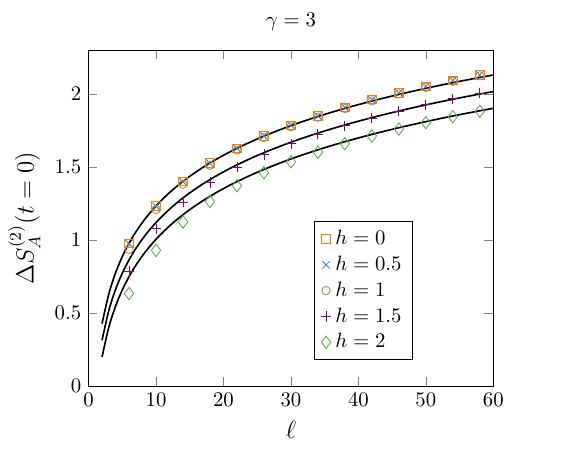}}
\caption{Rényi entanglement asymmetry $\Delta S_A^{(2)}$ as a function of the subsystem length $\ell$ for different values of $h$ and $\gamma$. The dots are the exact numerical values of the asymmetry computed using Eq.~\eqref{eq:numerics}. The solid lines correspond to the asymptotic result of Eq.~\eqref{eq:Deltaspn} while the dashed ones correspond to the evaluation of $\Delta S_A^{(2)}$ without the saddle point approximation in the Fourier transformation~\eqref{eq:FT} of the charged moments~\eqref{eq:Balpha_inftyt0}. }\label{fig:sp}
\end{figure}

\begin{figure}[t]
\centering
\subfigure
{\includegraphics[width=0.48\textwidth]{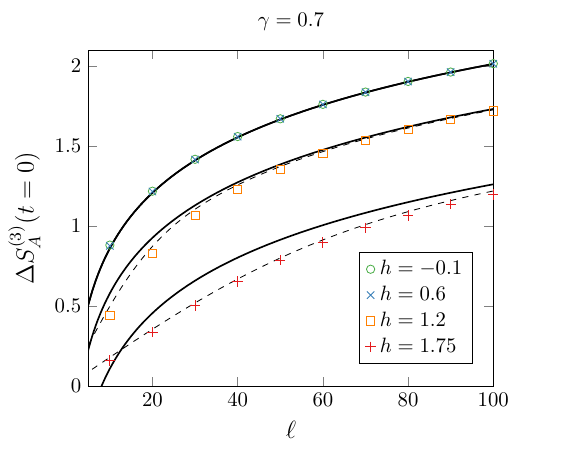}}
\subfigure
{\includegraphics[width=0.48\textwidth]{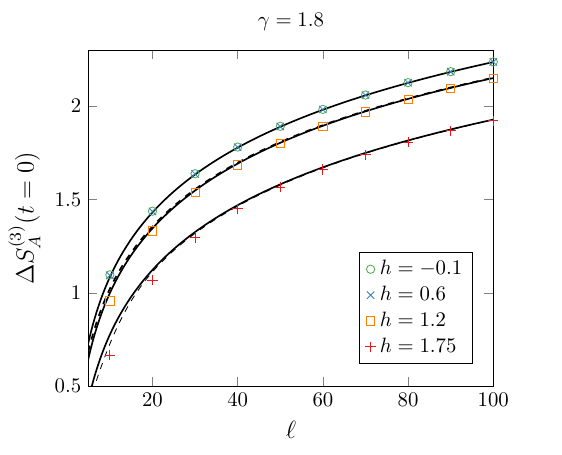}}
\caption{R\'enyi entanglement asymmetry $\Delta S_A^{(3)}$ for the ground state of the XY spin chain as a function of $\ell$ for different values of $h$ and $\gamma$. The points represent the exact numerical value obtained with Eq.~\eqref{eq:numerics}. The solid lines are the result of Eq.~\eqref{eq:Deltaspn} for large subsystem sizes while the dashed ones have been obtained calculating exactly the Fourier transformation~\eqref{eq:FT} of the charged moments $Z_n(\boldsymbol{\alpha})$ using their analytic expression~\eqref{eq:Balpha_inftyt0}.  }\label{fig:eqfactn32}
\end{figure}

In Figs.~\ref{fig:sp} and~\ref{fig:eqfactn32}, we investigate the validity of Eq.~\eqref{eq:Deltaspn} for $n=2$ and $n=3$ respectively. In these plots, we represent the ground state entanglement asymmetry as a function of the subsystem size taking different couplings $h$ and $\gamma$. The points are the exact numerical values of $\Delta S_A^{(n)}$ calculated with Eq.~\eqref{eq:numerics}. The dashed lines correspond to assume the prediction of Eq.~\eqref{eq:Balpha_inftyt0} for the charged moments
and then calculate numerically its exact Fourier transform~\eqref{eq:FT} to get $\Delta S_A^{(n)}$. In this case, we obtain a good agreement with the numerical points for all the values of $h$ and $\gamma$ considered. The solid lines represent the asymptotic behavior obtained in Eq.~\eqref{eq:spfactn} using the saddle point approximation. Observe that, for the range of subsystem sizes considered,  Eq.~\eqref{eq:Deltaspn} describes well the exact numerical results for $|h|\leq 1$ and any $\gamma$, both at $n=2$ and $n=3$. The same occurs for $|h|>1$ and $\gamma>1$. However, for $|h|>1$ and $\gamma<1$, the saddle point approximation requires to consider larger subsystems.

\begin{figure}[t]
\centering
\includegraphics[width=0.45\textwidth]{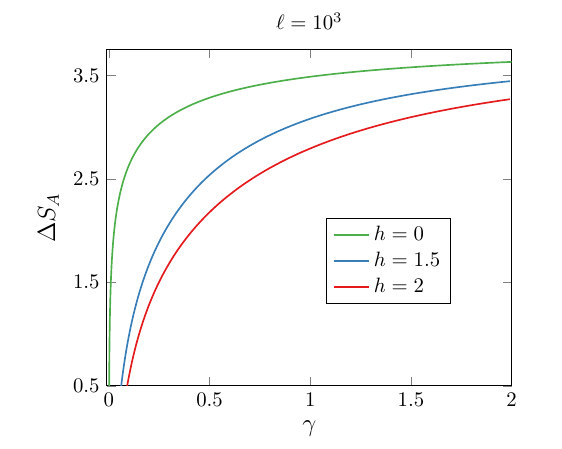}
\includegraphics[width=0.45\textwidth]{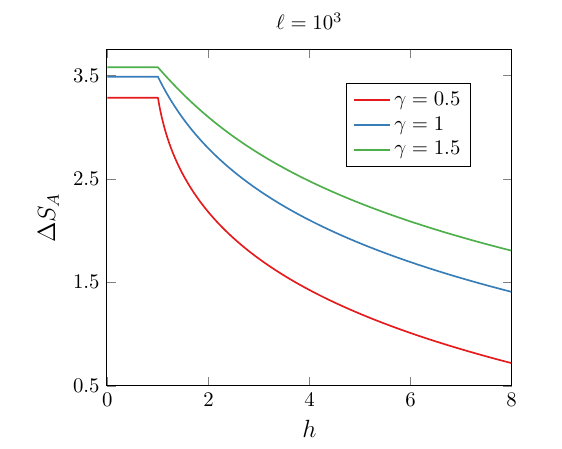}
\caption{Plot of the asymptotic expression~\eqref{eq:Deltaspn} in the limit $n\to1$ of the entanglement asymmetry for the ground state of the XY spin chain as a function of the anisotropy parameter $\gamma$ and several values of the external magnetic field $h$ (left panel) and viceversa (right panel). In both plots we take as subsystem length  $\ell=10^3$.}\label{fig:fgamma}
\end{figure}

In Fig.~\ref{fig:fgamma}, we plot the saddle point approximation of Eq.~\eqref{eq:Deltaspn} for $\Delta S_A$ as a function of $\gamma$ and several fixed values of $h$ (left panel) and viceversa (right panel) taking as susbsystem size $\ell=1000$ in both cases. Observe in the left panel that $\Delta S_A$ grows monotonically with the anisotropy parameter $\gamma$. Therefore, by varying $\gamma$, we can tune how much the $U(1)$ symmetry generated by $Q$ is broken. In particular, as we already pointed out, at $\gamma=0$, the Hamiltonian~\eqref{eq:Ham_XY} corresponds to the XX spin chain which commutes with $Q$. Hence the ground state respects the corresponding $U(1)$ symmetry and the entanglement asymmetry is expected to vanish. However, according to the asymptotic expression~\eqref{eq:Deltaspn}, $\Delta S_A\to-\infty$ when $\gamma\to 0$. The reason of this apparent discrepancy is that the limits $\ell\to \infty$ and $\gamma\to0$ do not commute. The other remarkable property of the ground state entanglement asymmetry can be seen in the right panel. As evident also from Eq.~\eqref{eq:g_gs_explicity}, for large $\ell$, the entanglement asymmetry is independent of the transverse magnetic field $h$ in the ferromagnetic phase ($|h|<1$) while, in the paramagnetic phase ($|h|>1$), it monotonically decreases with $h$. In fact, at $h\to\pm \infty$, the ground state of the XY spin chain is $\ket{\uparrow \uparrow \cdots \uparrow}$ and $\ket{\downarrow \downarrow \cdots \downarrow}$ respectively, which are eigenstates of $Q$, and $\Delta S_A^{(n)}=0$.  When we take this limit in the asymptotic expression~\eqref{eq:Deltaspn}, the entanglement asymmetry diverges  $\Delta S_A^{(n)}\to -\infty$ since the limits $\ell\to \infty$ and $h\to\pm\infty$ do not commute,  similarly to the case $\gamma\to 0$.

Finally, it is interesting to note that the asymptotic result~\eqref{eq:Deltaspn} for $\Delta S_A^{(n)}$ admits an interpretation in terms of the density of \textit{Cooper pairs} in the ground state $\ket{0}$. Observe that the factor $g(\gamma, h)$ that enters in Eq.~\eqref{eq:Deltaspn} only depends, as an integral in momentum space, on the quantity
$\sin^2\Delta_k$, see Eq.~\eqref{eq:g_gs}. Using the two-point correlation matrix $\Gamma$ of Eq.~\eqref{eq:Gammat0}, it is easy to see that $\sin\Delta_k$ is related to the correlator $\langle d_{2\pi-k}^\dagger d_k^\dagger\rangle$ by the equality $\langle d_{2\pi-k}^\dagger d_k^\dagger\rangle=i \sin\Delta_k/2$. The modulus $|\langle d_{2\pi-k}^\dagger d_k^\dagger\rangle|$ can be thought as the density of Cooper pairs of momentum $k$ that the state $\ket{0}$ contains. Therefore, since $\Delta S_A^{(n)}$ is proportional to the logarithm of $\ell g(\gamma, h)$ according to Eq.~\eqref{eq:Deltaspn}, it monotonically increases with the density of Cooper pairs present in the state $\ket{0}$ and the $U(1)$ symmetry associated to particle conservation is more broken. In fact, this symmetry is respected if and only if the correlations $\langle d_{2\pi-k}^\dagger d_{k}^\dagger\rangle$ vanish, i.e. in the absence of Cooper pairs. This interpretation of Cooper pairs as the excitations responsible of how much the particle number symmetry is broken will be further supported in the next section, where we elaborate a quasi-particle picture for the entanglement asymmetry after a quench in terms of them.

\section{Entanglement asymmetry out-of-equilibrium}\label{sec:quench}
In this section, we study the global quantum quench from the ground state of the XY spin chain~\eqref{eq:Ham_XY} with $\gamma\neq 0$, $\ket{\Psi(0)}=\ket{0}$, which breaks the particle number symmetry generated by $Q$, to the XX spin chain Hamiltonian $H_{\rm XX}$, which corresponds to take $\gamma=0$ and $h=0$ in Eq.~\eqref{eq:Ham_XY} and, therefore, it commutes with $Q$ and the $U(1)$ symmetry is expected to be dynamically restored in the subsystem $A$, i.e. $\lim_{t\to\infty}[\rho_A(t), Q_A]=0$. Thus the time-evolved state is
\begin{equation}\label{eq:quench}
 \ket{\Psi(t)}= e^{-i tH_{\rm XX}}\ket{\Psi(0)}.
\end{equation}
In order to evaluate the time evolution of the entanglement asymmetry in this quench protocol, we first derive a quasi-particle description for the dynamics of the charged moments defined in Eq.~\eqref{eq:Znalpha}.

\subsection{Time evolution of the charged moments}
 
  In Sec.~\ref{sec:xy_gs}, we have exploited the fact that the reduced density matrix $\rho_A$ of the ground state of the XY spin chain is Gaussian and, in virtue of Wick theorem, the charged moments $Z_n(\boldsymbol{\alpha}, t=0)$ are univocally determined by the two-point correlation matrix $\Gamma$ of Eq.~\eqref{eq:numerics}. Since the XX Hamiltonian is quadratic in terms of the fermionic operators $\boldsymbol{c}_j$, Eq.~\eqref{eq:numerics} also applies for the reduced density matrix $\rho_A(t)$ of subsystem $A$ after the quench. Furthermore, given that the post-quench Hamiltonian preserves the translational invariance of the system, the time-evolved two-point correlation matrix $\Gamma(t)$ is still block Toeplitz and reads~\cite{FC-08}
\begin{equation}
\Gamma_{jj'}(t)=\int_{-\pi}^{\pi}\frac{{\rm d}k}{2\pi}e^{-ik(j-j')} \mathcal{G}(k,t),\quad j,j'=1,\dots, \ell,
\end{equation}
where the symbol $\mathcal{G}(k,t)$ is now 
\begin{equation}\label{eq:symbol_time}
\mathcal{G}(k,t)=\left(\begin{array}{cc}
    \cos\Delta_k & -i e^{-2it\epsilon_{\rm XX}(k)}\sin\Delta_k\\
     ie^{2it\epsilon_{\rm XX}(k)}\sin\Delta_k& -\cos\Delta_k
    \end{array}\right),
\end{equation}
with $\cos \Delta_k$ and $\sin \Delta_k$ defined in Eqs.~\eqref{eq:cs} and $\epsilon_{\mathrm{XX}}(k)=-\cos(k)$ is the one-particle dispersion relation of the post-quench Hamiltonian $H_{\mathrm{XX}}$. 

In order to find the analytic expression that describes the charged moments $Z_n(\boldsymbol{\alpha}, t)$ in the ballistic regime $t,\ell\to \infty$ with $\zeta=t/\ell$ fixed, we first determine their stationary value at large times. It can be obtained by averaging the time dependent terms in the symbol $\mathcal{G}(k,t)$ of Eq.~\eqref{eq:symbol_time}. As $t\to \infty$, the terms $e^{\pm 2it\epsilon_{\rm XX}(k)}$ average to zero and the symbol reduces to
\begin{equation}\label{eq:symbol_asymptotic}
\mathcal{G}(k,t\to \infty)=\left(\begin{array}{cc}
    \cos\Delta_k & 0\\
   0& -\cos\Delta_k
    \end{array}\right).
\end{equation}
Observe that the correlators $\braket{\Psi(t)|c_jc_{j'}|\Psi(t)}$ and $\braket{\Psi(t)|c^{\dagger}_jc^{\dagger}_{j'}|\Psi(t)}$ vanish in the stationary regime. This is the first signature of the dynamical restoration of the particle number symmetry in the subsystem $A$.

For $n=2$, the stationary behavior of $Z_2(\alpha, t)$ can be determined by applying the conjecture of Eq.~\eqref{eq:conj_1}, as we did in Eq.~\eqref{eq:asymp_charged_mom_n_2}
for the charged moments of the ground state. In this case,
\begin{equation}\label{eq:asympt_charged}
\log  Z_2(\alpha, t\to \infty) \sim \frac{\ell}{2} \int_0^{2\pi} \frac{{\rm d}k}{2\pi}
\log\det\left[\frac{I+\mathcal{G}_{\alpha }(k,t\to\infty)\mathcal{G}_{-\alpha}(k,t\to\infty)}{2}\right],
\end{equation}
and, using the time-averaged symbol of Eq. 
\eqref{eq:symbol_asymptotic}, we find
\begin{equation}
\log Z_2(\alpha, t\to\infty) \sim \ell\int_0^{2\pi} \frac{{\rm d}k}{2\pi}
h_2\left(n(k)\right),
\end{equation}
where we have introduced
\begin{equation}
    h_n(x)=\log\left[x^n+(1-x)^n\right]
\end{equation}
and $n(k)\equiv\braket{\Psi(0)|d^{\dagger}_kd_k|\Psi(0)}=(1-\cos\Delta_k)/2$ is the density of occupied modes with momentum $k$. This result implies that $Z_2(\alpha, t\to \infty) \sim Z_2(0, t\to \infty)$; in fact, we recover the result predicted in Ref.~\cite{FC-08} for the stationary value of the entanglement entropy in this quench protocol.
 
For $n>2$, we cannot employ the conjecture of Eq.~\eqref{eq:conj_1} to derive the stationary value of $Z_n(\boldsymbol{\alpha}, t)$ at large times.
In general, the expression~\eqref{eq:numerics} for the charged moments does not simplify as for the case $n=2$, cf. Eq.~\eqref{eq:charged_mom_n_2}, and it contains the
inverse matrix $(I-\Gamma(t))^{-1}$. Nevertheless, in Eq.~\eqref{eq:conj_2} of Appendix~\ref{app:finite_temperature}, we report a formula that predicts the asymptotic behavior
of a determinant like the one in Eq.~\eqref{eq:numerics}, with a product of block Toeplitz matrices that also includes the inverse of block Toeplitz matrices. Since the time-averaged symbol $I-\mathcal{G}(k, t\to\infty)$ of the matrix $I-\Gamma(t)$ is invertible, we can directly apply Eq.~\eqref{eq:conj_2} to~\eqref{eq:numerics} in the large time limit,
\begin{equation}
 \log Z_n(\boldsymbol{\alpha}, t\to\infty)\sim 
 \frac{\ell}{2}\int_0^{2\pi} \frac{{\rm d}k}{2\pi}\log \det\left[\left(\frac{I-\mathcal{G}(k,t\to\infty)}{2}\right)^n
\left(I+\prod_{j=1}^n\mathcal{W}_j(k)\right)\right],\end{equation}
where 
$\mathcal{W}_j(k)=(I+\mathcal{G}(k,t\to\infty))(I-\mathcal{G}(k,t\to\infty))^{-1}e^{i\alpha_{j, j+1}\sigma_z}$. Using Eq.~\eqref{eq:symbol_asymptotic} and calculating
directly the determinant, we find
\begin{equation}\label{eq:z_n_t_stationary}
 \log Z_n(\boldsymbol{\alpha}, t\to\infty)\sim
 \ell \int_0^{2\pi}\frac{{\rm d}k}{2\pi} h_n(n(k)),
\end{equation}
that is, $Z_n(\boldsymbol{\alpha}, t\to\infty)\sim Z_n(\boldsymbol{0}, t\to\infty)$.

\begin{figure}
     \centering
     \includegraphics[width=0.45\textwidth]{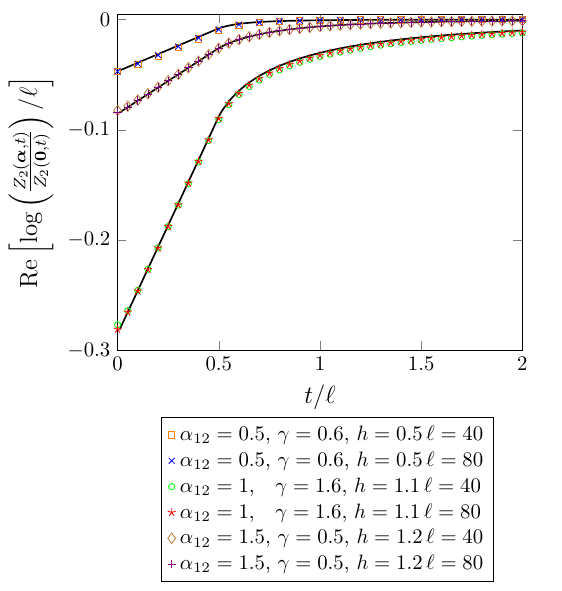}
     \includegraphics[width=0.45\textwidth]{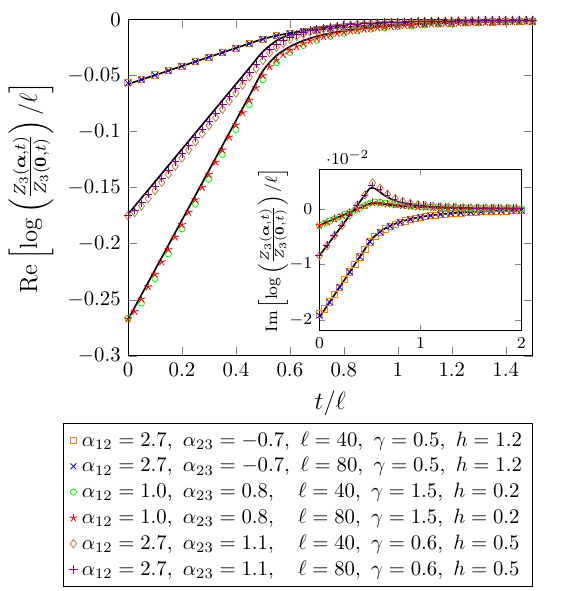}
     \caption{Time evolution of $Z_n(\boldsymbol{\alpha},0)$
      after the quench~\eqref{eq:quench} for $n=2$ (left panel) and $n=3$ (right panel). We plot it as a function of $t/\ell$ taking several initial ground states with different couplings $\gamma, h$ and various values of the subsystem size $\ell$ and the phases $\alpha_{j,j+1}$. The symbols were obtained numerically using Eq.~\eqref{eq:numerics} and the continuous lines correspond to the analytic prediction of Eq.~\eqref{eq:z_nevolution}.}
     \label{fig:moments}
\end{figure}

At this point, we know both the charged moments $Z_n(\boldsymbol{\alpha}, t)$ at
the initial time from Eq.~\eqref{eq:Balpha_inftyt0} and its asymptotic behavior at $t\to \infty$ in Eq.~\eqref{eq:z_n_t_stationary}. These two ingredients are enough to reconstruct the dynamics of $Z_n(\boldsymbol{\alpha}, t)$ for any finite time $t$ by
exploiting the quasi-particle picture of entanglement. The underlying idea is that the pre-quench initial state has very high
energy with respect to the ground state of the Hamiltonian governing the post-quench dynamics; hence, it can be seen as a source of quasi-particle excitations at $t=0$. We assume that quasi-particles are uniformly created in pairs with momenta $\pm k$ and
velocity $v(k)={\rm d} \epsilon_{\rm XX}(k)/{\rm d}k$. At a generic time $t$, the entanglement between a subsystem $A$ and $B$ is proportional to the total number of quasi-particles that were created at the same spatial point and are shared between $A$ and $B$ at that moment, which is given by the function $\min(2t|v(k)|, \ell)$. This idea has been firstly proposed to compute the entanglement dynamics after a global quantum quench in~\cite{cc-05, ac-17, ac-18}. However, we can also apply it here to determine the time evolution of the charged moments $Z_n(\boldsymbol{\alpha}, t)$, in the same way as it was done in Refs.~\cite{amc-22} and~\cite{amvc-23} for the tilted ferromagnetic and Néel states respectively. If we subtract from the stationary value~\eqref{eq:z_n_t_stationary} of $Z_n(\boldsymbol{\alpha}, t)$ its initial asymptotic behavior, obtained in Eq.~\eqref{eq:Balpha_inftyt0}, we get the contribution to $Z_n(\boldsymbol{\alpha}, t)$ at $t\to\infty$ of the pairs of entangled quasi-particle generated in the quench and shared between $A$ and $B$,
\begin{equation}\label{eq:stationary_charged_mom}
\log\left(\frac{Z_n(\boldsymbol{\alpha},t\to\infty)}{Z_n(\boldsymbol{\alpha}, t=0)}\right)\sim
\log Z_n(\boldsymbol{0}, t\to\infty) -\ell\int_0^{2\pi}\frac{{\rm d}k}{2\pi}\log \prod_{j=1}^nf_k(\alpha_{j,j+1}).
\end{equation}
This expression can be extended to finite times by properly counting the number of entangled excitations that $A$ and $B$ share at each moment. This can be done by simply inserting the function $\min(2\zeta |v(k)|, 1)$ in the momentum integrals of the right hand side of Eq.~\eqref{eq:stationary_charged_mom}.
We then obtain the exact time evolution after the quench of the charged moments~\eqref{eq:Znalpha} in the scaling limit $t, \ell \to \infty$ with $\zeta=t/\ell$ fixed,
\begin{equation}\label{eq:z_nevolution}
    Z_n(\boldsymbol{\alpha}, t)=Z_n(\boldsymbol{0}, t)e^{\ell(A_n(\boldsymbol{\alpha})+B_n(\boldsymbol{\alpha},\zeta))},
\end{equation}
where $Z_n(\boldsymbol{0},t)$ and $B_n(\boldsymbol{\alpha},\zeta)$ read respectively 
\begin{equation}
    \log Z_n(\boldsymbol{0},t)=\ell \int_0^{2\pi}\frac{{\rm d}k}{2\pi}\mathrm{min}(2\zeta |v(k)|,1) h_n(n(k))
\end{equation}
and
\begin{equation}\label{eq:Balpha}
    B_n(\boldsymbol{\alpha}, \zeta)=-\int_0^{2\pi}\frac{{\rm d} k}{4\pi}\mathrm{min}(2\zeta |v(k)|,1)\log\prod_{j=1}^n f_k(\alpha_{j,j+1}).
\end{equation}
The coefficient $ A_n(\boldsymbol{\alpha})$ is given in Eq.~\eqref{eq:A_n}. The expression~\eqref{eq:z_nevolution} is the main result of this section, and we benchmark it against exact numerical calculations in Fig.~\ref{fig:moments} taking as initial configuration the ground state of the XY spin chain for different values of the couplings $\gamma$ and $h$:
the symbols have been obtained using Eq.~\eqref{eq:numerics}, while the solid lines are Eq.~\eqref{eq:z_nevolution}. This expression is valid in the limit $\ell\to\infty$, and we observe that the agreement improves as $\ell$ increases.

\subsection{Time evolution of the entanglement asymmetry}
\begin{figure}
     \centering
     \includegraphics[width=0.485\textwidth]{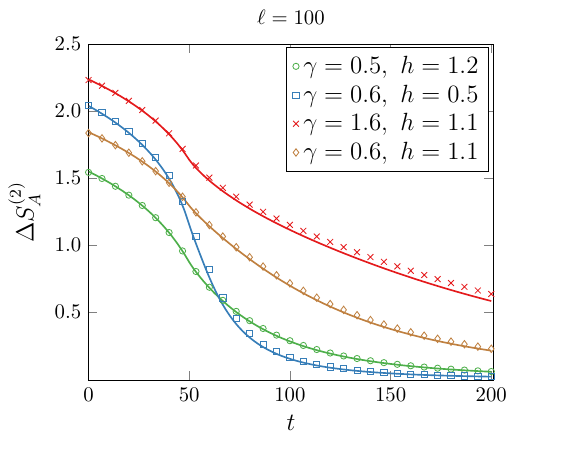}
     \includegraphics[width=0.495\textwidth]{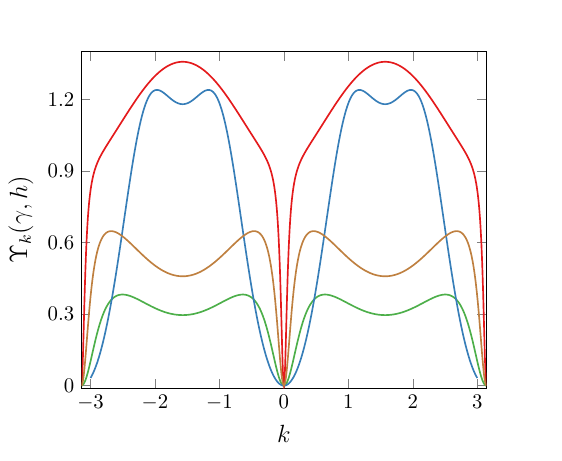}
     \includegraphics[width=0.485\textwidth]{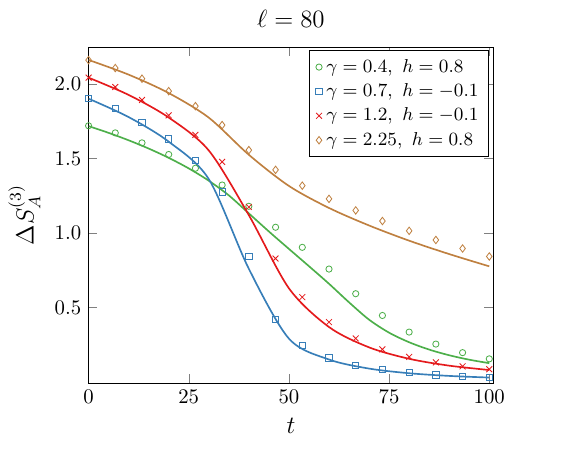}
     \includegraphics[width=0.485\textwidth]{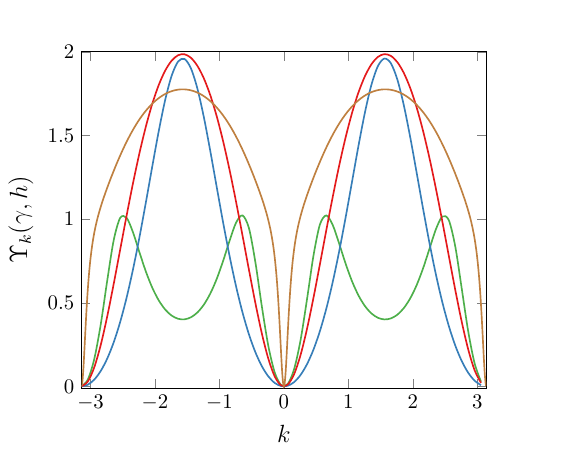}
     \caption{Left panels:  Time evolution of the R\'enyi entanglement asymmetry $\Delta S_{A}^{(n)}(t)$ after the quench~\eqref{eq:quench}. 
     The symbols are the exact numerical results for a subsystem of length $\ell=100$ ($n=2$) and $\ell=80$ ($n=3$), and different initial conditions for $\gamma,h$. 
     The continuous lines are our prediction obtained by plugging the charged moments reported in Eq.~\eqref{eq:z_nevolution} into Eqs.~\eqref{eq:FT} and~\eqref{eq:replicatrick}. 
    Right panels: Square of the density of the Cooper pairs at time $t=0$ in Eq.~\eqref{eq:cs}. The crossing of two densities is a necessary condition for the presence of quantum Mpemba effect, according to the criterion explained in the main text.}
     \label{fig:mpemba}
\end{figure}

We can now explicitly compute the time evolution of the entanglement asymmetry using the analytic result of the previous section. By plugging Eq.~\eqref{eq:z_nevolution} into Eq.~\eqref{eq:FT}, we obtain $\Delta S^{(n)}_A(t)$ in the scaling limit $t,\ell\to\infty$ with $\zeta$ fixed. We show the result in Fig.~\ref{fig:mpemba} for $n=2$ (left top panel) and $n=3$ (left bottom panel) and different choices of the parameters $\gamma$ and $h$ for the initial state. The agreement between our analytical prediction (solid lines) and the exact numerical computations (symbols) is overall very good in both cases, especially in the top panel, because the system size $\ell$ is bigger and also larger values of $n$ involve the computation of $n-1$-fold integral (according to Eq.~\eqref{eq:FT}), so bigger accuracy and precision.
Beyond the good matching, we remark that $\Delta S^{(n)}_A(t)$ tends to zero for large time (i.e. large $\zeta$). This is consistent with the fact that when we take the limit $t\to\infty$ in Eq.~\eqref{eq:z_nevolution}, the coefficient $B_n(\boldsymbol{\alpha})\to -A_n(\boldsymbol{\alpha})$ and, as we already saw, $Z_n(\boldsymbol{\alpha}, t\to\infty)\to Z_n(\boldsymbol{0}, t\to\infty)$. This implies that $\Delta S_A^{(n)}(t\to\infty)\to 0$ and the $U(1)$ symmetry is restored in subsystem $A$ in the stationary regime. This restoration was already observed in~\cite{amc-22}, see also Refs.~\cite{FCEC14, pvc-16}, for the quench from the tilted ferromagnetic state, which is the ground state of the XY spin chain along the curve $\gamma^2+h^2=1$. Another intriguing effect that we observe in Fig.~\ref{fig:mpemba} is that for some pairs of initial parameters, e.g. $\gamma=0.6$, $h=0.5$ and $\gamma=0.5$, $h=0.2$, the curves that 
the corresponding asymmetry $\Delta S_A^{(n)}(t)$ describes in time cross such that, for the state that initially breaks more the symmetry, the quench restores it earlier. This phenomenon was dubbed \textit{quantum Mpemba effect} in Ref.~\cite{amc-22}, which states that the more the system is initally out of equilibrium, the faster it relaxes. 
However, in the left panels of Fig.~\ref{fig:mpemba}, we can also see that this effect does not always occur. We can find pairs of initial couplings, e.g., $\gamma=0.6, h=0.5$ and $\gamma=1.6, h=1.2$, for which there is not a crossing between the curves and the symmetry is restored faster when the symmetry is less broken, i.e., for the smaller value of $\gamma$, $\gamma=0.6$. Let us investigate this phenomenon better to derive a condition under which we expect to observe the quantum Mpemba effect in the quenches~\eqref{eq:quench}.

Starting from Eq.~\eqref{eq:z_nevolution}, we aim to derive an effective closed-form approximation of $\Delta S^{(n)}_{A}(t)$ when the exponent in the charged moments $Z_n(\boldsymbol{\alpha}, t)$, $A_n(\boldsymbol{\alpha})+B_n(\boldsymbol{\alpha},\zeta)$ is small, i.e. for large values of time $t$. By using the Taylor expansion of an exponential function $e^{f(x)}$ when $f(x)\to 0$, the Fourier transform in Eq.~\eqref{eq:FT} can be performed analytically in that limit and we find 
\begin{equation}\label{eq:asympt}
\begin{split}
    \Delta S_{A}^{(n)}(t)&\simeq \frac{n \ell}{1-n}b(\zeta,\gamma,h),\\    
    b(\zeta,\gamma,h)&=\int_{-\pi}^{\pi} \frac{\mathrm{d}k}{4\pi}[1-\mathrm{min}(2\zeta |v(k)|,1)]\log\frac{1+\sqrt{1-\sin^2\Delta_k}}{2}.
    \end{split}
\end{equation}
This result represents the quasi-particle picture for the entanglement asymmetry in terms of Cooper pairs. As we discussed in Sec.~\ref{sec:xy_gs}, the term $\sin^2\Delta_k$ is identified with the density of Cooper pairs in the
initial state, i.e. $\sin^2\Delta_k=|\bra{\Psi(0)}d_{2\pi-k}^\dagger d_{k}^\dagger\ket{\Psi(0)}|^2$. Therefore, according to Eq.~\eqref{eq:asympt}, the entanglement asymmetry vanishes at large times as the number of Cooper pairs in the subsystem $A$ reduces ballistically to zero. This means that the rate at which the symmetry is restored is governed by the modes with the lowest group velocity $v(k)$. This observation is crucial to understand the occurrence of
the quantum Mpemba effect.

If we consider two different sets of couplings $h_1$, $\gamma_1$ and 
$h_2$, $\gamma_2$ for the initial ground state such that 
\begin{equation}\label{eq:mpemba_cond_1}
\Delta S_{A}^{(n)}(t=0,\gamma_1,h_1)<\Delta S_{A}^{(n)}(t=0,\gamma_2,h_2),
\end{equation}
then the quantum Mpemba effect occurs when there is a time, that we denote as $t_I$, after which the initial relation is inverted, i.e. 
\begin{equation}\label{eq:mpmeba_cond_2}
\Delta S_{A}^{(n)}(t,\gamma_1,h_1)>\Delta S_{A}^{(n)}(t,\gamma_2,h_2)\qquad \forall t>t_I.
\end{equation}
We can observe the quantum Mpemba effect if an only if conditions~\eqref{eq:mpemba_cond_1} and~\eqref{eq:mpmeba_cond_2} are satisfied.

Using the asymptotic expression~\eqref{eq:Deltaspn} for the ground state of the XY Hamiltonian, the condition~\eqref{eq:mpemba_cond_1} at $t=0$ can be rewritten in terms of the density of Cooper pairs of the two initial configurations as
\begin{equation}\label{eq:mpemba_cond_1_int}
\int_{-\pi}^\pi {\rm d}k\,\sin^2\Delta_k(\gamma_1, h_1)<\int_{-\pi}^\pi {\rm d}k\,\sin^2\Delta_k(\gamma_2, h_2).
\end{equation}

On the other hand, according to Eq.~\eqref{eq:asympt}, the inequality~\eqref{eq:mpmeba_cond_2} is satisfied if and only if $b(\zeta,\gamma_1,h_1)>b(\zeta,\gamma_2,h_2)$, for all $\zeta>\zeta_I=t_I/\ell$. 
It is clear that it is sufficient to enforce this second condition only for large times. Let us then study more carefully the behavior of the function $b(\zeta,\gamma_1,h_1)$ in the limit $t \to \infty$ or, equivalently, the limit $\zeta \to \infty$. In this case, it is useful to 
apply the identity,
\begin{equation}
    1-\mathrm{min}(2\zeta|v(k)|,1)=(1-2\zeta|v(k)|)\Theta\left(1-2\zeta|v(k)|\right),
\end{equation}
where $\Theta$ is the Heaviside Theta function, such that $\Theta(x)=1$ when $x>0$. Plugging this result in Eq.~\eqref{eq:asympt}, we firstly observe that
\begin{equation}
 b(\zeta,\gamma,h)= \int_{-\pi}^\pi \frac{{\rm d}k}{4\pi}(1-2\zeta|v(k)|)
 \Theta\left(1-2\zeta|v(k)|\right)\log\frac{1+\sqrt{1-\sin^2\Delta_k}}{2},
\end{equation}
is non-vanishing for the modes $-\zeta^{-1}<2v(k)<\zeta^{-1}$. At large times, since 
$|v(k)|=|\sin(k)|$, this condition is satisfied if 
$k^*(\zeta)=\arcsin{(1/(2\zeta))}$ exists such that
\begin{multline}\label{eq:b_kstar}
b(\zeta,\gamma,h) = \int_{-k^*(\zeta)}^{k^*(\zeta)} \frac{{\rm d}k}{4\pi}(1-2\zeta|v(k)|)\log\frac{1+\sqrt{1-\sin^2\Delta_k}}{2}\\
+\int_{\pi-k^*(\zeta)}^{\pi+k^*(\zeta)}\frac{{\rm d}k}{4\pi}(1-2\zeta|v(k)|)\log\frac{1+\sqrt{1-\sin^2\Delta_k}}{2}.
\end{multline}
Outside the critical lines $|h|\neq 1$, $\sin^2\Delta_k$ vanishes around $k=0$ and $\pi$ and, therefore, we 
can take the approximation $\log[(1+\sqrt{1-x})/2]\sim -x/4$,
\begin{equation}\label{eq:app_b_sinDelta}
 b(\zeta,\gamma,h)\simeq -\int_{-k^*(\zeta)}^{k^*(\zeta)} \frac{{\rm d}k}{16\pi}(1-2\zeta|v(k)|)\sin^2\Delta_k
 -\int_{\pi-k^*(\zeta)}^{\pi+k^*(\zeta)} \frac{{\rm d}k}{16\pi} (1-2\zeta|v(k)|)\sin^2\Delta_k.
\end{equation}
If we perform the change of variables $k'=k-\pi$ in the second integral of the expression above, we then find
\begin{equation}
     b(\zeta,\gamma,h)\simeq -\int_{-k^*(\zeta)}^{k^*(\zeta)} \frac{\mathrm{d}k}{16\pi}(1-2\zeta|v(k)|)\Upsilon_k(\gamma, h),
\end{equation}
where $\Upsilon_k(\gamma, h)=\sin^2\Delta_k(\gamma,h)+\sin^2\Delta_k (\gamma,-h)$.

Therefore, the condition~\eqref{eq:mpmeba_cond_2}, i.e. $b(\zeta, h_1, \gamma_1)>b(\zeta, h_2, \gamma_2)$ for large $\zeta$, to observe the quantum Mpemba effect can be re-expressed in terms of the densities of Cooper pairs in the initial states as
\begin{equation}\label{eq:mpemba_cond_2_int}
\int_{-k^*(\zeta)}^{k^*(\zeta)}{\rm d}k\, \Upsilon_k(\gamma_1, h_1)> 
\int_{-k^*(\zeta)}^{k^*(\zeta)}{\rm d}k\, \Upsilon_k(\gamma_2, h_2) \qquad {\rm for} \; \zeta>\zeta_I.
\end{equation}
Given the form of $\sin^2\Delta_k$, $\Upsilon_k(\gamma, h)$ is a definite positive, even function of $k$ that vanishes at $k=0$ for any value of $\gamma> 0$ and $|h|\neq 1$. Therefore, there always exists a large enough time $t_I$ for which the integral condition of Eq.~\eqref{eq:mpemba_cond_2_int} can be replaced by
\begin{equation}\label{eq:mpemba}  
\Upsilon_k(\gamma_1,h_1)>\Upsilon_k(\gamma_2, h_2), \quad k\in \left[-\arcsin\left({\frac{\ell}{2t_I}}\right),\arcsin\left({\frac{\ell}{2t_I}}\right)\right].
\end{equation}
Eqs.~\eqref{eq:mpemba_cond_1_int} and~\eqref{eq:mpemba} are the necessary and sufficient microscopic conditions to observe the quantum Mpemba effect between a pair of ground states of the XY spin chain after a quench to the XX spin chain. According to them, the quantum Mpemba effect occurs when the state that initially breaks less the symmetry, and therefore contains a smaller net number of Cooper pairs (condition~\eqref{eq:mpemba_cond_1_int}), has instead a larger density of Cooper pairs around the modes with the slowest velocity $v(k)$ (condition~\eqref{eq:mpemba}), which correspond to the momenta $k=0$ and $k=\pi$. This is a very natural condition since the entanglement asymmetry satisfies the quasi-particle picture of Eq.~\eqref{eq:asympt} and, therefore, its leading behavior at large times is determined by the slowest excitations. In the right panels of Fig.~\ref{fig:mpemba}, we plot the function $\Upsilon_k(\gamma, h)$ that enters in the condition~\eqref{eq:mpemba} for some of the initial states studied in the left panels of that figure: observe that, whenever the inequality~\eqref{eq:mpemba} is met for a pair of couplings $(\gamma, h)$ that also satisfy~\eqref{eq:mpemba_cond_1_int}, the curves that describe their asymmetries intersect at certain time and  Eq.~\eqref{eq:mpmeba_cond_2} is fulfilled. 
Notice that the simultaneous validity of Eqs.~\eqref{eq:mpemba_cond_1} and ~\eqref{eq:mpmeba_cond_2} then requires that the density of Cooper pairs corresponding to two different quenches should cross, as made explicit in Fig. \ref{fig:mpemba}. 
In addition, we observe that the conditions~\eqref{eq:mpemba_cond_1_int} and~\eqref{eq:mpemba_cond_2_int} are valid for any value of the Rényi index $n$. For the condition at $t=0$, the reason is that all the dependence on $\gamma$ and $h$ in Eq.~\eqref{eq:Deltaspn} is in the term $g(\gamma, h)$, which is independent of $n$. For the large time condition, the starting point~\eqref{eq:asympt} from which it is derived does not depend on $n$.

Many of the former considerations are valid generically in integrable systems \cite{abckmr-23}. Specializing on our quench, we can obtain a set of conditions for the quantum Mpemba effect equivalent to the microscopic ones but only involving the couplings $\gamma$, $h$ of the initial states. For the inequality~\eqref{eq:mpemba_cond_1} at $t=0$, this can be straightforwardly done using the asymptotic expression~\eqref{eq:Deltaspn}, together with Eq.~\eqref{eq:g_gs_explicity} for the term $g(\gamma, h)$. In the case of the condition~\eqref{eq:mpmeba_cond_2} at large times, we need to determine explicitly the leading behavior of $\Delta S_A^{(n)}(t)$ when $t\to\infty$. For $|h|\neq 1$, this can be done from Eq.~\eqref{eq:app_b_sinDelta} by expanding the functions $v(k)$ and $\sin^2\Delta_k$ around 
$k=0$ or $k=\pi$ in each integral and $k^*(\zeta)$ around $\zeta=\infty$. We find that, at leading order in large $\zeta$,
\begin{equation}
    \Delta S_{A}^{(n)}(t)= \frac{n}{384\pi(n-1)}\frac{ \gamma ^2 \left(h^2+1\right)}{\left(h^2-1\right)^2}  \frac{\ell}{\zeta^3}\,,
    \label{eq:exp2}
\end{equation}
i.e. it vanishes for large times as $t^{-3}$ for any value of $\gamma$ and $h$. The fact that the prefactor in Eq.~\eqref{eq:exp2} monotonically increases as a function of $\gamma$ and it depends non-trivially on $h$ reflects that it is not enough starting from a state with larger $\gamma$ to reach before $\Delta S_{A}^{(n)}(t)\to 0$, but the dependence on $h$ is crucial to observe the Mpemba effect. Fixing $\gamma$, we notice that, for $|h|<1$, Eq.~\eqref{eq:exp2} is a monotonically increasing function of $h$; since the initial asymmetry grows with $\gamma$ and does not depend on $h$ in this region, then it is necessary that $\gamma_2>\gamma_1$ and $h_2<h_1$ to satisfy the Mpemba conditions~\eqref{eq:mpemba_cond_1} and~\eqref{eq:mpmeba_cond_2}. In particular, they are always met by any pair of ground states with couplings belonging to the curve $h^2+x\gamma^2=1$ for a fixed parameter $x>0$, which describes an ellipse in the $(h,\gamma)$-plane. In fact, for any initial state on this curve,
\begin{equation}\label{eq:asytcube}
    \Delta S_{A}^{(n)}(t)= \frac{n}{384\pi(n-1)x}\left( \frac{ 2}{x\gamma^2}-1\right)  \frac{\ell}{\zeta^3}\,.
\end{equation}
In this case, the prefactor of the $t^{-3}$ decay is a monotonously decreasing function of $\gamma$, and, therefore, we always observe that the more the symmetry is broken, the faster it is restored. Interestingly, for large subsystems, the spectrum of the correlation matrix $\Gamma$ is the same for all the ground states along a curve $h^2+x\gamma^2=1$ and, consequently, they have equal entanglement entropy~\cite{fijk-07, aefq-16, aefq-17}. On the other hand, the discussion in the region $|h|>1$ is more involved because both the initial entanglement asymmetry~\eqref{eq:Deltaspn} and its large time behavior~\eqref{eq:exp2} are monotonic decreasing functions of $h$.
\begin{figure}[t]
     \centering
     \includegraphics[width=0.6\textwidth]{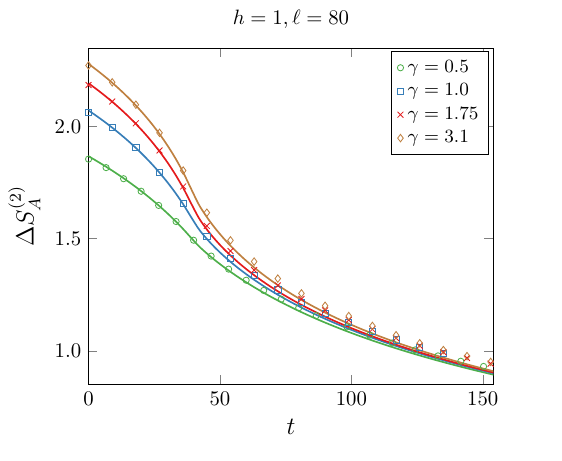}     
     \caption{ Time evolution of the R\'enyi entanglement asymmetry $\Delta S_{A}^{(2)}(t)$ after the quench~\eqref{eq:quench} from a critical state, $h=1$ and different $\gamma$'s. 
     The symbols are the exact numerical results for a subsystem of length  $\ell=80$. 
     The continuous lines are our prediction obtained by plugging the charged moments reported in Eq.~\eqref{eq:z_nevolution} into Eqs.~\eqref{eq:FT} and~\eqref{eq:replicatrick}. As explained in the main text, whatever is the initial value of $\gamma$, for large time $t$ the lines collapse and, eventually, the symmetry is restored almost simultaneously, a weak version of Mpemba effect.}
     \label{fig:critical}
\end{figure}

The replica limit $n\to 1$ in Eq.~\eqref{eq:exp2} is not well defined. In Appendix \ref{app:proof}, we carefully perform it, starting from the Fourier transform of the charged moments in Eq.~\eqref{eq:FT}. The final result reads
\begin{equation}\label{eq:asymptn1}
     \Delta S_A(t)=-\frac{\gamma ^2 \left(h^2+1\right) \ell}{384 \pi  \left(h^2-1\right)^2\zeta^3}\log\left[\frac{\gamma ^2 \left(h^2+1\right) \ell}{384 \pi  \left(h^2-1\right)^2\zeta^3}\right].
\end{equation}
Observe that, while the Rényi entanglement asymmetry in Eq.~\eqref{eq:exp2} decays to zero at large times as $\ell^4/t^3$, in the limit $n\to 1$ it behaves as $\ell^4\log(t)/t^3$, being the logarithmic correction $\log(t)$ a particular feature of this case. This also happens for the von Neumann entanglement entropy, as it has been found in \cite{FC-08}.

When $|h|=1$, we can find an expression similar to Eq.~\eqref{eq:exp2}. In this case, $\sin^2\Delta_{k}\neq 0$ at $k=0$ and the 
approximation of Eq.~\eqref{eq:app_b_sinDelta} is not valid. If we take Eq.~\eqref{eq:b_kstar} instead and we expand at leading order the 
integrands around the modes $k=0$ and $k=\pi$ respectively and the function $k^*(\zeta)$ around $\zeta=\infty$, then we obtain
\begin{equation}
    \Delta S_{A}^{(n)}(t)= \frac{n}{n-1}\frac{\log 2}{8\pi}\frac{\ell}{\zeta}\,.
    \label{eq:exp3}
\end{equation}
We observe that the behavior of the entanglement asymmetry as a function of $\zeta$ is different if the initial configuration is the ground state of a critical Hamiltonian or not: in the former case, it decreases as $1/\zeta$, while if we start outside the critical line the decay to zero is algebraically faster, as $1/\zeta^3$. Therefore, if we consider a critical state and a non-critical one that breaks more the symmetry, the symmetry is always restored faster in the latter. This can be seen as a {\it strong} quantum Mpemba effect. In fact, in the classical Mpemba effect, the system relaxes exponentially to the equilibrium state, but in certain particular situations, the decay is exponentially faster, a phenomenon dubbed as strong Mpemba effect~\cite{raz2}. By analogy, in our quantum setup, the asymmetry reaches the equilibrium always following a power law but with a smaller exponent in the case of critical states, so in a much slower fashion.
In addition, note that the prefactor of Eq.~\eqref{eq:exp3} does not depend on $\gamma$, while the initial entanglement asymmetry along the lines $|h|=1$ grows monotonically with $\gamma$ according to Eq.~\eqref{eq:Deltaspn}. This means that, independently of how much the symmetry is initially broken, for critical states, it is restored (almost) at the same time, as we show in Fig.~\ref{fig:critical}.
We can call this phenomenon {\it weak} quantum Mpemba effect.

As occurs in the non-critical region, the limit $n\to1$ in Eq.~\eqref{eq:exp3} is also not well defined. Repeating the same steps as in Appendix \ref{app:proof} for the gapped phase, we find 
\begin{equation}\label{eq:asympcritc}
     \Delta S_A(t)=-\frac{\ell\log 2}{8 \pi\zeta}\log\left[\frac{\ell }{8 \pi\zeta}\right],
\end{equation}
which differs from the result of Eq.~\eqref{eq:exp3} for the Rényi entanglement asymmetry in the logarithmic correction $\log(t)$.

\section{Conclusions}\label{sec:conclusions}
In this manuscript, we have investigated the $U(1)$ symmetry breaking in the XY spin chain using the entanglement asymmetry, completing the analysis initiated in Ref.~\cite{amc-22} for the tilted ferromagnetic state and specializing the general discussion on the quantum Mpemba effect for integrable systems done in Ref.\cite{abckmr-23}. We have first studied the behavior of the entanglement asymmetry in the ground state of this model, finding that, at leading order, it grows logarithmically with the subsystem size whether ($|h|\neq 1$) or not the system is gapped. We remark that this is quite different with respect to what happens for the total entanglement entropy, quantity from which the entanglement asymmetry is defined: when $|h|\neq 1$, the entanglement entropy saturates to a constant value for large subsystems~\cite{peschel-xy, ijk-05}, while a violation of the area law occurs only along the critical lines, where the entropy scales logarithmically with the subsystem size~\cite{cc-04}. Another important result of this work is that we find that the entanglement asymmetry depends on the density of the Cooper pairs, $|\langle d_{2\pi-k}d_k\rangle|$, of the ground state. This is a natural result if we take into account that the breaking of the $U(1)$ particle number symmetry in the XY spin chain can be traced back to the presence of superconducting pairing terms in the corresponding fermionic Hamiltonian. 

In addition, we have investigated the evolution of the entanglement asymmetry in a global quantum quench, starting from the ground state of the XY spin chain and letting the system evolve with the XX Hamiltonian that preserves the particle number such that the symmetry is dynamically restored in the subsystem. With the help of the quasi-particle picture of entanglement, we have derived a closed-form analytic expression for the asymmetry at large times, from which we have deduced the necessary and sufficient conditions to observe the quantum Mpemba effect in terms of the density of Cooper pairs of the initial
states. Essentially, if the density of the slowest Cooper pairs is larger for the state that breaks less the symmetry, then the Mpemba physics shows up, meaning that the more the symmetry is broken, the faster it is restored. The set of microscopic conditions that we obtain here are in agreement with the criteria derived in Ref.~\cite{abckmr-23} for an arbitrary integrable quantum system. 

It would be interesting to investigate several questions in the future. The first one is an explanation of the mechanism of the quantum Mpemba effect when the symmetry is restored by a non-integrable Hamiltonian, or the evolution is non-unitary. This analysis has been initiated in Ref.~\cite{amc-22} for systems of a few sites, showing the robustness of this phenomenon also if the evolution Hamiltonian is non-integrable. So far, only the breaking of Abelian symmetries has been investigated, but we would like to use the entanglement asymmetry to explore the symmetry breaking of non-Abelian groups. Finally, the analysis done in this manuscript has revealed that the critical lines of the XY model, $|h|=1$, are peculiar since an extra term appears in the charged moments. It would be interesting to find its exact expression and determine its (subleading) contribution to the entanglement asymmetry, not only to have a more accurate prediction of it but also to understand if it contains information about the underlying conformal field theory that describes these critical lines. 

\section*{Acknowledgments} 
We thank Bruno Bertini, Katja Klobas and Colin Rylands for useful discussions and collaboration on a related topic \cite{abckmr-23}. 
PC and FA acknowledge support from ERC under Consolidator grant number 771536 (NEMO). SM thanks support from Caltech Institute for Quantum Information and Matter and the Walter Burke Institute for Theoretical Physics at Caltech. The work of IK was supported in part by the NSF grant DMR-1918207.

\begin{appendices}

\section*{Appendices}

\section{Derivation of the asympotic behavior of the ground state charged moments for integer $n>2$}\label{app:finite_temperature}

In this Appendix, we show how to obtain the expression in Eq.~\eqref{eq:Balpha_inftyt0}
for the charged moments $Z_n(\boldsymbol{\alpha})$ in the ground state
of the XY spin chain. Observe that, in general, for $n>2$  the inverse matrix $(I-\Gamma)^{-1}$ cannot be removed from Eq.~\eqref{eq:numerics} as we did in Eq.~\eqref{eq:charged_mom_n_2} when $n=2$. In general, the inverse of a block Toeplitz matrix is not block Toeplitz and the result of Eq.~\eqref{eq:conj_1} cannot be in principle 
applied. However, in Ref.~\cite{amvc-23}, we found a corollary of Eq.~\eqref{eq:conj_1} for the determinant of a product of Toeplitz matrices that involves as well the inverse of block Toeplitz matrices. According to it, if we further include in the determinant of Eq.~\eqref{eq:conj_1} the inverse of the block Toeplitz matrices $T_\ell[g_j']$, then for large $\ell$,
\begin{equation}\label{eq:conj_2}
\det\left(I+\prod_{j=1}^n T_\ell[g_j]T_\ell[g_j']^{-1}\right)\sim
e^{\ell A'},
\end{equation}
where
\begin{equation}
A'=\int_0^{2\pi}\frac{{\rm d}k}{2\pi} \log\det\left[I+\prod_{j=1}^n g_j(k)g_j'(k)^{-1}\right].
\end{equation}

However, observe that the symbol of the matrix $I-\Gamma$ is $I-\mathcal{G}$, with $\mathcal{G}$ given by Eq.~\eqref{eq:gs_symbol}. This symbol is not invertible and Eq.~\eqref{eq:conj_2} cannot be applied. We can bypass
this issue by considering the system at finite temperature $1/\beta$ and then take the limit $\beta\to \infty$. In fact, the state of the spin chain at temperature $1/\beta$ is described by the Gibbs ensemble $\rho_\beta=e^{-\beta H}/Z$, where $Z={\rm Tr}(e^{-\beta H})$. The two-point correlation function $\Gamma_\beta$ associated to $\rho_\beta$ is block Toeplitz with symbol
\begin{equation}
\mathcal{G}_\beta(k)=\tanh\left(\frac{\beta \epsilon(k)}{2}\right)
\left(\begin{array}{cc}
    \cos\Delta_k & -i\sin\Delta_k \\
    i\sin\Delta_k & -\cos\Delta_k
    \end{array}\right),
\end{equation}
where $\epsilon(k)$ is the one-particle dispersion relation of the XY spin chain Hamiltonian. Observe that, in the zero temperature limit $\beta\to \infty$, $\mathcal{G}_\beta(k)$ yields the ground state symbol $\mathcal{G}(k)$ reported in Eq.~\eqref{eq:gs_symbol}. The advantage of $\mathcal{G}_\beta$ is that $I-\mathcal{G}_\beta$ is invertible and Eq.~\eqref{eq:conj_2} can be applied to determine the asymptotic behavior of the 
charged moments $Z_n(\boldsymbol{\alpha}, \beta)$ at finite temperature and large subsystem size $\ell$. We find
\begin{equation}
 Z_n(\boldsymbol{\alpha},\beta)=e^{\ell A_n(\boldsymbol{\alpha}, \beta)},
\end{equation}
with
\begin{equation}\label{eq:A_n_beta}
A_n(\boldsymbol{\alpha},\beta)=
\int_{0}^{2\pi} \frac{{\rm d}k}{4\pi} 
\log\det\left[\left(\frac{I-\mathcal{G}_\beta(k)}{2}\right)^n 
\left(I+\prod_{j=1}^n\mathcal{W}_{\beta, j}(k)\right)\right],
\end{equation}
where $\mathcal{W}_{\beta,j}(k)$ stands for the $2\times 2$ matrix
$\mathcal{W}_{\beta, j}(k)=(I+\mathcal{G}_\beta(k))(I-\mathcal{G}_\beta(k))^{-1}e^{i\alpha_{jj+1}\sigma_z}$. 
If we now consider the quotient
\begin{equation}
\frac{Z_n(\boldsymbol{\alpha}, \beta)}{Z_n(\boldsymbol{0}, \beta)}=
e^{\ell\left[A_n(\boldsymbol{\alpha},\beta)-A_n(\boldsymbol{0}, \beta)\right]},
\end{equation}
and take the limit $\beta\to\infty$, we find Eq.~\eqref{eq:Balpha_inftyt0} with
\begin{equation}
A_n(\boldsymbol{\alpha})=\lim_{\beta\to\infty}\left[A_n(\boldsymbol{\alpha},\beta)-A_n(\boldsymbol{0}, \beta)\right].
\end{equation}
By calculating explicitly the determinant in the integrand of Eq.~\eqref{eq:A_n_beta} 
for different integer values of $n$, one can check that this limit 
actually yields the factorized form of Eq.~\eqref{eq:A_n} for the coefficient $A_n(\boldsymbol{\alpha})$.

\section{Large time behavior of the von Neumann entanglement asymmetry}\label{app:proof}
In Eq.~\eqref{eq:exp2}, we notice that the replica limit $n\to 1$ is not well-defined. Therefore, in this Appendix, we carefully derive the asymptotic expression in the limit $t\to\infty$ of the entanglement asymmetry~\eqref{eq:def}. 
As already observed in Ref.~\cite{abckmr-23}, in this regime we can explicitly compute the Fourier transform in Eq. \eqref{eq:FT}.
Indeed, by expanding the charged moments for small values of $(1-\mathrm{min}(2\zeta |v(k)|,1))$ (i.e. large $\zeta$), we find  
\begin{equation}
\begin{split}
   \frac{ \mathrm{Tr}(\rho_{A, Q}^n)}{ \mathrm{Tr}(\rho_{A}^n)}=&\int_{-\pi}^\pi \frac{{\rm d}\alpha_1\dots{\rm d}\alpha_n}{(2\pi)^n}\exp\left[\ell\int_{-\pi}^{\pi}\frac{{\rm d} k}{4\pi}(1-\mathrm{min}(2\zeta |v(k)|,1))\log\prod_{j=1}^n f_k(\alpha_{j,j+1})\right]\\
   &\simeq 1+\ell \int_{-\pi}^\pi \frac{{\rm d}\alpha_1\dots{\rm d}\alpha_n}{(2\pi)^n}\int_{-\pi}^{\pi}\frac{{\rm d} k}{4\pi}(1-\mathrm{min}(2\zeta |v(k)|,1))\log\prod_{j=1}^n f_k(\alpha_{j,j+1}).
\end{split}
\end{equation}
The integral above has been done in Eq. [sm-54] of \cite{abckmr-23}, and, by identifying $\max[\vartheta(k), 1-\vartheta(k)]=(1+\sqrt{1-\sin^2\Delta_k})/2$ and $\min[\vartheta(k), 1-\vartheta(k)]=(1-\sqrt{1-\sin^2\Delta_k})/2$, we can report here the final result in our case,
\begin{equation}
\begin{split}
    \frac{ \mathrm{Tr}(\rho_{A, Q}^n)}{ \mathrm{Tr}(\rho_{A}^n)}\simeq &\left(1+\ell \int_{-\pi}^{\pi}\frac{{\rm d} k}{4\pi}(1-\mathrm{min}(2\zeta |v(k)|,1))\log\frac{1+\sqrt{1-\sin^2\Delta_k}}{2}\right)^n\\
&+\sum_{j=-\infty}^{-1}\left[(-1)^j\frac{\ell}{j} \int_{-\pi}^{\pi}\frac{{\rm d} k}{4\pi}(1-\mathrm{min}(2\zeta |v(k)|,1))\left(\frac{1+\sqrt{1-\sin^2\Delta_k}}{1-\sqrt{1-\sin^2\Delta_k}}\right)^j\right]^n.
    \end{split}
\end{equation}
We can deduce the replica limit $n\to1$ after doing an analytic continuation of the result above to any complex value of $n$ and, in the large time regime, we find
\begin{equation}\label{eq:asyn1}
\begin{split}
    \Delta S_A(t)&=-\lim_{n\to 1}\partial_n \frac{ \mathrm{Tr}(\rho_{A, Q}^n)}{ \mathrm{Tr}(\rho_{A}^n)} \simeq\\
   &-\sum_{j=-\infty}^{-1}(-1)^j\left[\frac{\ell}{j} \int_{-\pi}^{\pi}\frac{{\rm d} k}{4\pi}(1-\mathrm{min}(2\zeta |v(k)|,1))\left(\frac{1+\sqrt{1-\sin^2\Delta_k}}{1-\sqrt{1-\sin^2\Delta_k}}\right)^{j}\right.\\ & \times \left.\log\left((-1)^j\frac{\ell}{j} \int_{-\pi}^{\pi}\frac{{\rm d} k}{4\pi}(1-\mathrm{min}(2\zeta |v(k)|,1))\left(\frac{1+\sqrt{1-\sin^2\Delta_k}}{1-\sqrt{1-\sin^2\Delta_k}}\right)^{j}\right)\right].
 \end{split}
\end{equation}
If $t$ is sufficiently large, then $(1-\mathrm{min}(2\zeta |v(k)|,1))$ becomes zero everywhere, except for a finite interval around the points $k=0,\pi$ where the magnitude of the velocity is minimal. Therefore, since we are interested in the leading order behavior in $\zeta$, we can restrict the sum in Eq.~\eqref{eq:asyn1} to $j=-1$. Moreover, by expanding $k^*(\zeta)\sim 1/(2\zeta)$ for large $\zeta$ and $\sin^2\Delta_k$ around $k=0, \pi$, we obtain for $|h|\neq 1$
\begin{equation}
    \int_{-k^*(\zeta)}^{k^*(\zeta)}\frac{{\rm d} k}{16\pi}(1-2|k|\zeta)\Upsilon_{k\simeq 0}(\gamma, h)=\frac{\gamma ^2 \left(h^2+1\right) }{384 \pi  \left(h^2-1\right)^2\zeta^3},
\end{equation}
and, finally, we get Eq. \eqref{eq:asymptn1} of the main text.

Along the critical lines $|h|=1$, close to $k=0$, we find 
\begin{equation}
    \frac{1+\sqrt{1-\sin^2\Delta_k}}{1-\sqrt{1-\sin^2\Delta_k}}=1+O(k).
\end{equation}
Therefore, the main difference with respect to the non-critical case is that the leading term at large $\zeta$ in the series of Eq.~\eqref{eq:asyn1} is not $j=-1$ but we have now to consider all of them. By taking into account that $\sum_{j=-\infty}^{-1}(-1)^j/j=\log 2$, we obtain at leading order in $\zeta$
\begin{equation}
    \Delta S_A(t) \simeq
   -\ell\,\log(2) \left[\int_{-k^*(\zeta)}^{k^*(\zeta)}\frac{{\rm d} k}{4\pi}(1-2|k|\zeta )\right]\times\log\left[\ell \int_{-k^*(\zeta)}^{k^*(\zeta)}\frac{{\rm d} k}{4\pi}(1-2|k|\zeta)\right],
\end{equation}
from which Eq.~\eqref{eq:asympcritc} is derived.

\section{Comparison between the charged moments $Z_n(\boldsymbol{\alpha})$ and the FCS}

The expression for the charged moments in Eq.~\eqref{eq:Znalpha} when $n=1$ is also known as full counting statistics (FCS), $\chi(\alpha)=\mathrm{Tr}(\rho_A e^{i\alpha Q_A})$, see Refs.~\cite{Cherng, ia-13, stephan, gec-18, ARV21} for different studies of it in the XY spin chain. Given the result for generic $n$ in Eq.~\eqref{eq:Balpha_inftyt0} for the ground state, one might be tempted to deduce that, if the $U(1)$ symmetry is broken, the charged moments $Z_n(\boldsymbol{\alpha})$ factorize into the product of the FCS with different phases $\alpha_{j,j+1}$. However, using the results for the FCS obtained in~\cite{Cherng, ARV21}, we will show in the following that this is not always true.

The FCS can be cast as the determinant of a Toeplitz 
matrix with symbol $f(e^{i\Delta_k}, \alpha/2)$~\cite{Cherng, ARV21}, where the function $f$ is given in Eq.~\eqref{eq:f}.
Thus one can use the theorems on the asymptotic behavior of Toeplitz 
determinants to analyze $\chi(\alpha)$ for $\ell\gg 1$.
 For $|h|>1$ and
any value of $\alpha$ or when $h<1$ and $\alpha\in(-\pi/2, \pi/2)$,
the symbol $f(e^{i\Delta_k}, \alpha/2)$ is a non-zero continuous function in $k$ and
the Szeg\H o theorem holds,
\begin{equation}\label{eq:fcs1}
 \log \chi(\alpha)\sim \ell \int_0^{2\pi}\frac{{\rm d}k}{2\pi}\log f\left(e^{i\Delta_k}, \frac{\alpha}{2}\right).
\end{equation}
We observe that the integral satisfies the following equality 
\begin{equation}\label{eq:fcs3}
 \int_0^{2\pi}\frac{{\rm d}k}{2\pi}\log f\left(e^{i\Delta_k}, \frac{\alpha}{2}\right)= \int_0^{2\pi}\frac{{\rm d}k}{4\pi}\log f\left(\cos(i\Delta_k), \alpha\right),
\end{equation}
which implies that, in this regime of the parameters, the result in Eq.~\eqref{eq:Balpha_inftyt0} is a factorization of the charged moments into the FCS.
However, when $|h| < 1$ and $\alpha \in [-\pi,-\pi/2] \cup [\pi/2, \pi] $, the symbol $f(e^{i\Delta_k}, \alpha)$ 
acquires winding number $+1$. In this case, the prediction in Eq.~\eqref{eq:fcs1} is not valid and 
it must be modified as
\begin{equation}\label{eq:fcs_winding}
\log \chi(\alpha)\sim \ell \left(\int_0^{2\pi}\frac{{\rm d}k}{2\pi}\log [e^{-ik}f\left(e^{i\Delta_k}, \frac{\alpha}{2}\right)] +\log (-z_0)\right).
\end{equation}
If we consider the analytic continuation of $f(e^{i\Delta_k}, \alpha)$ from the unit circle $z=e^{ik}$ to
the complex plane, then $z_0$ denotes the zero of such analytic continuation with $|z_0| < 1$ and 
closest to the unit circle $z=e^{ik}$. This point can be either
\begin{equation}
 z_0=\frac{h+\sqrt{h^2+\gamma^2-1}}{1+\gamma},\quad \mbox{or}\quad
 z_0=\frac{h+\sqrt{h^2+\gamma^2\cos^2(\alpha)-1}}{1-\gamma\cos(\alpha)}.
\end{equation}
The presence of this winding number is the responsible that the charged moments $Z_n(\boldsymbol{\alpha})$ do not {\it exactly} factorize when $\ell\to \infty$ into the FCS $\mathrm{Tr} (\rho_A e^{i\alpha_{j,j+1}Q_A})$. In other words, 
if we compare Eq.~\eqref{eq:Balpha_inftyt0} with~\eqref{eq:fcs1} and
\eqref{eq:fcs_winding}, the factorization only works in principle when $\alpha_{j,j+1}\in(-\pi/2, \pi/2)$ for all $j$. But taking into account the periodicity properties in $\alpha_{j, j+1}$
of the charged moments, it can be extended to $\alpha_{j,j+1}\in[-\pi,\pi]$ by introducing the parameter $\sigma_j$, which vanishes if $|\alpha_{j,j+1}| \leq \pi/2$ and  $\sigma_j=\pi$ otherwise, i.e. we can write
\begin{equation}\label{eq:uno}
    Z_n(\boldsymbol{\alpha})\sim  Z_n(\boldsymbol{0})\prod_{j=1}^n e^{i\sigma_j/2} \mathrm{Tr}(\rho_A e^{i(\alpha_{j,j+1} -\sigma_j)Q_A}).
\end{equation}
The term $\sigma_j=\pi$ ensures that we are always in the regime where Eq.~\eqref{eq:fcs1} is valid.

The Fourier transform of the FCS yields the probability distribution
$p(q)$ for the transverse magnetization $Q_A$ (or particle number) to take the value $q$.
We can make a comparison between our final result in Eq.~\eqref{eq:Deltaspn} and the R\'enyi-Shannon entropy for the
distribution $p(q$), or R\'enyi number entropy,   
\begin{equation}
 H_n=\frac{1}{1-n}\log \sum_q p(q)^n,   
\end{equation}
where $p(q)$ is the probability for the observable $Q_A$ to take the value $q$.
The result for $H_n$ reads
\begin{equation}
    H_n=\frac{1}{2}\log \ell+O(1),
\end{equation}
where the $O(1)$ term does depend on $(\gamma,h)$ and, in general, it is different with respect to what we find in Eq.~\eqref{eq:Deltaspn}. In fact, it is clear from that expression that the entanglement asymmetry only takes into account the number of Cooper pairs as the $O(1)$ term only depends on $\sin\Delta_k$, and not on the total number of fermions which contribute to $H_n$.

\end{appendices}


\begin{thebibliography}{10}
\bibitem{cool}
E. B. Mpemba and D. G. Osborne, \textit{Cool?}, \href{https://iopscience.iop.org/article/10.1088/0031-9120/4/3/312}{Phys. Educ. \textbf{4}, 172
(1969).}

\bibitem{clathrate}
Y. H. Ahn, H. Kang, D. Y. Koh, and H. Lee, \textit{Experimental verifications of Mpemba-like behaviors of clathrate hydrates,} \href{https://link.springer.com/article/10.1007/s11814-016-0029-2}{Korean
Jour. of Chem. Engin. \textbf{33}, 1903 (2016)}.

\bibitem{polymers}
C. Hu, J. Li, S. Huang, H. Li, C. Luo, J. Chen, S. Jiang, and L. An, 
\textit{Conformation Directed Mpemba Effect on Polylactide Crystallization,} 
\href{https://doi.org/10.1021/acs.cgd.8b01250}{Cryst. Growth Des. {\bf 18}, 5757 (2018)}.

\bibitem{magneto}
P. Chaddah, S. Dash, K. Kumar, and A. Banerjee, \textit{Overtaking while approaching equilibrium,} \href{
https://doi.org/10.48550/arXiv.1011.3598
}{arXiv:1011.3598}

\bibitem{carbon}
P. A. Greaney, G. Lani, G. Cicero, and J. C. Grossman, 
\textit{Mpemba-Like Behavior in Carbon Nanotube Resonators,} \href{https://link.springer.com/article/10.1007/s11661-011-0843-4}{Metal. and Mat. Trans. A \textbf{42}, 3907 (2011).}

\bibitem{granular}
A. Lasanta, F. Vega Reyes, A. Prados, and A. Santos, \textit{When the Hotter Cools More Quickly: Mpemba Effect in Granular Fluids,}
\href{https://journals.aps.org/prl/abstract/10.1103/PhysRevLett.119.148001}{Phys. Rev. Lett. \textbf{119}, 148001 (2017).}

\bibitem{cold}
T. Keller, V. Torggler, S. B. J\"ager, S. Sch\"tz, H. Ritsch and G. Morigi, \textit{Quenches across the self-organization transition in multimode cavities,} \href{https://iopscience.iop.org/article/10.1088/1367-2630/aaa161}{New J. Phys. \textbf{20}, 025004 (2018).}


\bibitem{raz}
Z. Lu and O. Raz, \textit{Nonequilibrium thermodynamics of the Markovian Mpemba effect and its inverse,} \href{https://www.ncbi.nlm.nih.gov/pmc/articles/PMC5441807/}{PNAS \textbf{114}, 5083 (2017)}.

\bibitem{raz2}
I. Klich, O. Raz, O. Hirschberg, and M. Vucelja, {\it The Mpemba index and anomalous relaxation,} \href{
https://doi.org/10.1103/PhysRevX.9.021060}{Phys.
Rev. X {\bf 9}, 021060 (2019).}


\bibitem{kumar}
A. Kumar and J. Bechhoefer, \textit{
 Exponentially faster cooling in a colloidal system,} \href{https://www.nature.com/articles/s41586-020-2560-x}{
Nature \textbf{584}, 64 (2020).}

\bibitem{wv-22}
M. R. Walker and M. Vucelja, \textit{Mpemba effect in terms of mean first passage time,} \href{https://arxiv.org/abs/2212.07496}{	arXiv:2212.07496 (2022).}

\bibitem{teza2023relaxation}
G. Teza, R. Yaacobu and O. Raz,
\textit{Relaxation shortcuts through boundary coupling,} 
\href{https://doi.org/10.1103/PhysRevLett.131.017101}{Phys. Rev. Lett. \textbf{131}, 017101 (2023)}.

\bibitem{wbv-23}
M. R. Walker, S. Bera, and M. Vucelja,
{\it Optimal transport and anomalous thermal relaxations},
\href{https://doi.org/10.48550/arXiv.2307.16103}{arXiv:2307.16103}.

\bibitem{bwv-23}
S. Bera, M. R. Walker, and M. Vucelja
{\it Effect of dynamics on anomalous thermal relaxations and information exchange},
\href{https://doi.org/10.48550/arXiv.2308.04557}{arXiv:2308.04557}.

\bibitem{quantum1}
A. Nava and M. Fabrizio, \textit{Lindblad dissipative dynamics in the presence of phase coexistence,} \href{https://journals.aps.org/prb/abstract/10.1103/PhysRevB.100.125102}{Phys. Rev. B \textbf{100}, 125102 (2019).}

\bibitem{quantum2}
S. Kochsiek, F. Carollo, and I. Lesanovsky, \textit{Accelerating the approach of dissipative quantum spin systems towards stationarity through global spin rotations,} \href{https://journals.aps.org/pra/abstract/10.1103/PhysRevA.106.012207}{Phys. Rev. A \textbf{106}, 012207 (2022).}

\bibitem{quantum3}
F. Carollo, A. Lasanta, and I. Lesanovsky, \textit{Exponentially Accelerated Approach to Stationarity in Markovian Open Quantum Systems through the Mpemba Effect,}
\href{https://journals.aps.org/prl/abstract/10.1103/PhysRevLett.127.060401}{Phys. Rev. Lett. \textbf{127}, 060401 (2021).}

\bibitem{quantum4}
S. K. Manikandan, \textit{Equidistant quenches in few-level quantum systems,}
\href{https://journals.aps.org/prresearch/abstract/10.1103/PhysRevResearch.3.043108}{Phys. Rev. Research \textbf{3}, 043108 (2021).}

\bibitem{quantum5}
F. Ivander, N. Anto-Sztrikacs, and D. Segal, \textit{Hyper-acceleration of quantum thermalization dynamics by bypassing long-lived coherences: An analytical treatment,} \href{https://doi.org/10.1103/PhysRevE.108.014130}{Phys. Rev. E {\bf 108}, 014130 (2023).}

\bibitem{quantum6}
A. K. Chatterjee, S. Takada, and H. Hayakawa, \textit{Quantum Mpemba effect in a quantum dot with reservoirs,} 
\href{https://doi.org/10.1103/PhysRevLett.131.080402}{Phys. Rev. Lett. {\bf 131}, 080402 (2023).}

\bibitem{amc-22}
F. Ares, S. Murciano, and P. Calabrese,
{\it Entanglement asymmetry as a probe of symmetry breaking},
\href{https://doi.org/10.1038/s41467-023-37747-8}{Nature Communications {\bf 14}, 2036 (2023)}.

\bibitem{amvc-23} 
F. Ares, S. Murciano, E. Vernier, and P. Calabrese,
{\it Lack of symmetry restoration after a quantum quench: an entanglement asymmetry study},
\href{https://doi.org/10.21468/SciPostPhys.15.3.089}{SciPost Phys. {\bf 15}, 089 (2023)}.

\bibitem{abckmr-23} 
C. Rylands, K. Klobas, F. Ares, P. Calabrese, S. Murciano, and B. Bertini,
{\it Microscopic origin of the quantum Mpemba effect in integrable systems},
\href{https://doi.org/10.48550/arXiv.2310.04419}{arXiv:2310.04419}.

\bibitem{bkcccr-23}
B. Bertini, K. Klobas, M. Collura, P. Calabrese, and C. Rylands,
{\it Dynamics of charge fluctuations from asymmetric initial states},
\href{https://doi.org/10.48550/arXiv.2306.12404}{arXiv:2306.12404}.

\bibitem{exp-ti}
L. Kh. Joshi, J. Franke, A. Rath, F. Ares, S. Murciano, F. Kranzl, R. Blatt, P. Zoller, B. Vermersch, P. Calabrese, C. F. Roos, and M. Joshi, 
{\it Observing the quantum Mpemba effect in quantum simulations}, 
\href{https://arxiv.org/abs/2401.04270}{arXiv:2401.04270}.

\bibitem{fac-23}
F. Ferro, F. Ares, and P. Calabrese, {\it Non-equilibrium entanglement asymmetry for discrete groups: the example of the XY spin chain},
\href{http://arxiv.org/abs/2307.06902}{arXiv:2307.06902}.

\bibitem{cm-23}
L. Capizzi and M. Mazzoni,
{\it Entanglement asymmetry in the ordered phase of many-body systems: the Ising Field Theory},
\href{https://doi.org/10.1007/JHEP12(2023)144}{JHEP {\bf 12} (2023) 144}.

\bibitem{cv-23}
L. Capizzi and V. Vitale,
{\it A universal formula for the entanglement asymmetry of matrix product states},
\href{https://doi.org/10.48550/arXiv.2310.01962}{arXiv:2310.01962}.

\bibitem{klich08scaling}
I. Klich and L. S. Levitov, {\it Scaling of entanglement entropy and superselection rules},
\href{https://arxiv.org/abs/0812.0006}{arXiv:0812.0006}.

\bibitem{lr-14}
N. Laflorencie and S. Rachel, 
{\it Spin-resolved entanglement spectroscopy of critical spin chains and Luttinger liquids},
\href{http://dx.doi.org/10.1088/1742-5468/2014/11/P11013}{J. Stat. Mech. (2014) P11013}.

\bibitem{goldstein}
M. Goldstein and E. Sela, 
{\it Symmetry-Resolved Entanglement in Many-Body Systems},
\href{http://dx.doi.org/10.1103/PhysRevLett.120.200602}{Phys. Rev. Lett. {\bf 120}, 200602 (2018)}.

\bibitem{xavier}
J. C. Xavier, F. C. Alcaraz, and G. Sierra, 
{\it Equipartition of the entanglement entropy}, 
\href{https://journals.aps.org/prb/abstract/10.1103/PhysRevB.98.041106}{Phys. Rev. B {\bf  98}, 041106 (2018)}.

\bibitem{cgs-18}
E. Cornfeld, M. Goldstein, and E. Sela,
{\it Imbalance Entanglement: Symmetry Decomposition of Negativity},
\href{https://doi.org/10.1103/PhysRevA.98.032302}{Phys. Rev. A {\bf 98}, 032302 (2018)}.

\bibitem{mbc-21}
S. Murciano, R. Bonsignori, and P. Calabrese
{\it Symmetry decomposition of negativity of massless free fermions},
\href{https://doi.org/10.21468/SciPostPhys.10.5.111}{SciPost Phys. {\bf 10}, 111 (2021)}.

\bibitem{czzc-21}
L. Capizzi and P. Calabrese,
{\it Symmetry resolved relative entropies and distances in
conformal field theory},
\href{https://doi.org/10.1007/JHEP10(2021)195}{JHEP {\bf 10} (2021) 195}.

\bibitem{dge-23}
G. Di Giulio and J. Erdmenger,
{\it Symmetry-resolved modular correlation functions in free fermionic theories}
\href{https://doi.org/10.1007/JHEP07%282023%29058}{JHEP {\bf 07} (2023) 058}.

\bibitem{brc-19}
R. Bonsignori, P. Ruggiero, and P. Calabrese, 
{\it Symmetry resolved entanglement in free fermionic systems}, 
\href{https://doi.org/10.1088/1751-8121/ab4b77}{J. Phys. A {\bf 52}, 475302 (2019)}.

\bibitem{mgc-20}
S. Murciano, G. Di Giulio, and P. Calabrese, 
{\it Entanglement and symmetry resolution in two dimensional free quantum field theories}, 
\href{https://link.springer.com/article/10.1007/JHEP08(2020)073}{JHEP {\bf 08} (2020) 073}.

\bibitem{pbc-21-1}
G. Parez, R. Bonsignori, and P. Calabrese, 
{\it Quasiparticle dynamics of symmetry resolved
entanglement after a quench: the examples of conformal field theories and free fermions},
\href{https://doi.org/10.1103/PhysRevB.103.L041104}{Phys. Rev. B {\bf 103}, L041104 (2021)}.

\bibitem{pbc-21-2}
G. Parez, R. Bonsignori, and P. Calabrese, 
{\it Exact quench dynamics of symmetry resolved
entanglement in a free fermion chain}, 
\href{https://doi.org/10.1088/1742-5468/ac21d7}{J. Stat. Mech. (2021) 093102}.

\bibitem{pvcc-22}
L. Piroli, E. Vernier, M. Collura, and P. Calabrese, 
{\it Thermodynamic symmetry resolved entanglement entropies in integrable systems}, 
\href{https://doi.org/10.1088/1742-5468/ac7a2d}{J. Stat. Mech. (2022) 073102}.

\bibitem{bcckr-22}
B. Bertini, P. Calabrese, M. Collura, K. Klobas, and C. Rylands, 
{\it Nonequilibrium Full Counting Statistics and Symmetry-Resolved Entanglement from Space-Time Duality},
\href{https://doi.org/10.1103/PhysRevLett.131.140401}{Phys. Rev. Lett. {\bf 131}, 140401 (2023)}.

\bibitem{mac-23}
S. Murciano, V. Alba, and P. Calabrese,
{\it Symmetry-resolved entanglement in fermionic systems with dissipation}
\href{https://doi.org/10.1088/1742-5468/ad0224} {J. Stat. Mech. (2023) 113102}.

\bibitem{lukin}
A. Lukin, M. Rispoli, R. Schittko, M. E. Tai, A. M. Kaufman, S. Choi, V. Khemani, J. Leonard, and M. Greiner, 
{\it Probing entanglement in a many-body localized system}, 
\href{https://dx.doi.org/10.1126/science.aau0818}{Science {\bf 364}, 6437 (2019)}.

\bibitem{azses}
D. Azses, R. Haenel, Y. Naveh, R. Raussendorf, E. Sela, and E. G. Dalla Torre, 
{\it Identification of Symmetry-Protected Topological States on Noisy Quantum Computers}, 
\href{https://doi.org/10.1103/PhysRevLett.125.120502}{Phys. Rev. Lett. {\bf 125}, 120502 (2020)}.

\bibitem{neven}
A. Neven, J. Carrasco, V. Vitale, C. Kokail, A. Elben, M. Dalmonte, P. Calabrese, P. Zoller, B. Vermersch, R. Kueng, and B. Kraus, 
{\it Symmetry-resolved entanglement detection using partial transpose moments}, 
\href{https://doi.org/10.1038/s41534-021-00487-y}{npj Quantum Info. {\bf 7}, 1 (2021)}.

\bibitem{vitale}
V. Vitale, A. Elben, R. Kueng, A. Neven, J. Carrasco, B. Kraus, P. Zoller, P. Calabrese, B. Vermersch, and M. Dalmonte, 
{\it Symmetry-resolved dynamical purification in synthetic
quantum matter}, 
\href{https://doi.org/10.21468/SciPostPhys.12.3.106 }{SciPost Phys. 12, 106 (2022)}.

\bibitem{rath}
A. Rath, V. Vitale, S. Murciano, M. Votto, J. Dubail, R. Kueng, C. Branciard, P. Calabrese, and B. Vermersch, 
{\it Entanglement barrier and its symmetry resolution: theory and experiment}, 
\href{https://doi.org/10.1103/PRXQuantum.4.010318}{PRX Quantum {\bf 4}, 010318 (2023)}.

\bibitem{mhms-22}
Z. Ma, C. Han, Y. Meir, and E. Sela, 
{\it Symmetric inseparability and number entanglement
in charge conserving mixed states}, 
\href{https://doi.org/10.1103/PhysRevA.105.042416}{Phys. Rev. A {\bf 105}, 042416 (2022)}.

\bibitem{hlw-94}
C. Holzhey, F. Larsen, and F. Wilczek, 
{\it Geometric and renormalized entropy in conformal field theory}, \href{http://dx.doi.org/10.1016/0550-3213(94)90402-2}{Nucl. Phys. B {\bf 424}, 443 (1994)}.

\bibitem{cc-04}
P. Calabrese and J. Cardy, 
{\it Entanglement entropy and quantum field theory}, 
\href{http://dx.doi.org/10.1088/1742-5468/2004/06/P06002}{J. Stat. Mech. (2004) P06002}.

\bibitem{vermersch}
B. Vermersch, A. Elben, L. M. Sieberer, N. Y. Yao, and P. Zoller, 
{\it Probing scrambling using statistical correlations between randomized measurements}, 
\href{https://doi.org/10.1103/PhysRevX.9.021061}{Phys. Rev. X {\bf 9}, 021061 (2019)}.

\bibitem{brydges}
T. Brydges, A. Elben, P. Jurcevic, B. Vermersch, C. Maier, B. P. Lanyon, P. Zoller, R. Blatt, and C. F. Roos, {\it Probing entanglement entropy via randomized measurements},
\href{http://dx.doi.org/10.1126/science.aau4963}{Science {\bf 364}, 260 (2019)}.

\bibitem{huang}
H.-Y. Huang, R. Kueng, and J. Preskill, 
{\it Predicting many properties of a quantum system
from very few measurements}, 
\href{https://doi.org/10.1038/s41567-020-0932-7}{Nature Phys. {\bf 16}, 1050 (2020)}.

\bibitem{elben}
A. Elben, S. T. Flammia, H.-Y. Huang, R. Kueng, J. Preskill, B. Vermersch, and P. Zoller, 
{\it The randomized measurement toolbox}, 
\href{https://doi.org/10.1038/s42254-022-00535-2}{Nat. Rev. Phys. {\bf 5}, 9 (2023)}.

\bibitem{hms-23}
C. Han, Y. Meir, and E. Sela, 
{\it Realistic Protocol to Measure Entanglement at Finite Temperatures},
\href{https://doi.org/10.1103/PhysRevLett.130.136201}{Phys. Rev. Lett. {\bf 130}, 136201 (2023)}.

\bibitem{ktm-82}
J. Kurmann, H. Thomas, and G. Müller, 
{\it Antiferromagnetic long-range order in the
anisotropic quantum spin chain}, 
\href{https://doi.org/10.1016/0378-4371(82)90217-5}{Physica A {\bf 112}, 235 (1982)}.

\bibitem{ms-85}
G. Müller and R.E. Shrock, 
{\it Implications of direct-product ground states in the one-
dimensional quantum XYZ and XY spin chains}, 
\href{https://doi.org/10.1103/PhysRevB.32.5845}{Phys. Rev. B {\bf 32}, 5845 (1985)}.

\bibitem{lieb}
E. Lieb, T. Schultz, and D. Mattis, 
{\it Two soluble models of an antiferromagnetic chain}, 
\href{https://doi.org/10.1016/0003-4916(61)90115-4}{Ann. Phys. {\bf 16}, 407 (1961)}.


\bibitem{p-03}
I. Peschel, 
\textit{Calculation of reduced density matrices from correlation functions},
\href{https://iopscience.iop.org/article/10.1088/0305-4470/36/14/101}{J. Phys. A {\bf 36}, L205 (2003)}.

\bibitem{FC-08}
M. Fagotti and P. Calabrese,
{\it Evolution of entanglement entropy following a quantum quench: Analytic results for the XY chain in a transverse magnetic field}, 
\href{https://doi.org/10.1103/PhysRevA.78.010306}{Phys. Rev. A {\bf 78}, 010306(R) (2008)}.


\bibitem{FC10} 
M. Fagotti and P. Calabrese,
\textit{Entanglement entropy of two disjoint blocks in XY chains},
\href{https://doi.org/10.1088/1742-5468/2010/04/P04016}{J. Stat. Mech. (2010) P04016}.

\bibitem{balian}
R. Balian and E. Brezin, 
{\it Nonunitary Bogoliubov transformations and extension of Wick’s theorem}, 
\href{https://link.springer.com/article/10.1007/BF02710281}{Il Nuovo Cimento B {\bf 64}, 37 (1969)}.

\bibitem{cc-05} 
P. Calabrese and J. Cardy,
\textit{Evolution of Entanglement Entropy in One-Dimensional Systems},
\href{https://doi.org/10.1088/1742-5468/2005/04/P04010}{J. Stat. Mech. (2005) P04010}.

\bibitem{ac-17} 
V. Alba and P. Calabrese,
\textit{Entanglement and thermodynamics after a quantum quench in integrable systems},
\href{https://doi.org/10.1073/pnas.1703516114}{PNAS {\bf 114}, 7947 (2017)}.

\bibitem{ac-18} 
V. Alba and P. Calabrese, 
\textit{Entanglement dynamics after quantum quenches in generic integrable systems}, 
\href{https://scipost.org/10.21468/SciPostPhys.4.3.017}{SciPost Phys. {\bf 4}, 017 (2018)}.

\bibitem{FCEC14} 
M. Fagotti, M. Collura, F. H. L. Essler, P. Calabrese,
\textit{Relaxation after quantum quenches in the spin-1/2 Heisenberg XXZ chain},
\href{https://doi.org/10.1103/PhysRevB.89.125101}{Phys. Rev. B {\bf 89}, 125101 (2014)}.

\bibitem{pvc-16}
L. Piroli, E. Vernier, and P. Calabrese, 
{\it Exact steady states for quantum quenches in integrable Heisenberg spin chains}, 
\href{http://dx.doi.org/10.1103/PhysRevB.94.054313}{Phys. Rev. B {\bf 94}, 054313 (2016)}.

\bibitem{fijk-07}
F. Franchini, A. R. Its, B.-Q. Jin, and V. E. Korepin
{\it Ellipses of Constant Entropy in the XY Spin Chain},
\href{https://doi.org/10.1088/1751-8113/40/29/019}{J. Phys. A: Math. Theor. {\bf 40}, 8467 (2007)}.

\bibitem{aefq-16}
F. Ares, J. G. Esteve, F. Falceto, and A. R. De Queiroz,
{\it On the M\"obius transformation in the entanglement entropy of fermionic chains},
\href{http://dx.doi.org/10.1088/1742-5468/2016/04/043106}{J. Stat. Mech. (2016) 043106}.

\bibitem{aefq-17}
F. Ares, J. G. Esteve, F. Falceto, and A. R. De Queiroz,
{\it Entanglement entropy and Möbius transformations for critical fermionic chains},
\href{https://doi.org/10.1088/1742-5468/aa71dc}{J. Stat. Mech. (2017) 063104}.

\bibitem{peschel-xy}
I. Peschel,
{\it On the entanglement entropy for a XY spin chain},
\href{ https://doi.org/10.1088/1742-5468/2004/12/P12005}{J. Stat. Mech. (2004) P12005}.

\bibitem{ijk-05}
A. R. Its, B.-Q. Jin, and V. E. Korepin,
{\it Entanglement in XY Spin Chain},
\href{https://doi.org/10.1088/0305-4470/38/13/011}{J. Phys. A: Math. Gen. {\bf 38}, 2975 (2005)}.

\bibitem{Cherng} 
R. W. Cherng and E. Demler, 
{\it Quantum Noise Analysis of Spin Systems Realized with Cold Atoms}, 
\href{https://doi.org/10.1088/1367-2630/9/1/007}{New J. Phys. {\bf 9}, 7 (2007)}.

\bibitem{ia-13}
D. A. Ivanov and A. G. Abanov, 
{\it Characterizing correlations with full counting statistics: classical Ising and quantum XY spin chains}, 
\href{https://doi.org/10.1103/PhysRevE.87.022114}{Phys. Rev. E {\bf 87}, 022114 (2013)}.

\bibitem{stephan}
J.-M. Stéphan, 
{\it Emptiness formation probability, Toeplitz determinants, and conformal field theory}, 
\href{https://doi.org/10.1088/1742-5468/2014/05/P05010}{J. Stat. Mech. (2014) P05010}.

\bibitem{gec-18}
S. Groha, F. H. L. Essler, and P. Calabrese, 
{\it Full Counting Statistics in the Transverse Field Ising Chain}, \href{https://scipost.org/SciPostPhys.4.6.043}{SciPost Phys. {\bf 4}, 043 (2018)}.

\bibitem{ARV21}
F. Ares, M. A. Rajabpour, and J. Viti, 
{\it Exact full counting statistics for the staggered magnetization and the domain walls in the XY spin chain},
\href{https://doi.org/10.1103/PhysRevE.103.042107}{Phys. Rev. E {\bf 103}, 042107 (2021)}.

\end{thebibliography}
\end{document}